\RequirePackage[2020-02-02]{latexrelease}

\documentclass[11pt]{article}
\usepackage[margin=1in]{geometry}
\usepackage{amsmath}
\usepackage{amssymb}
\usepackage{amsthm}
\usepackage{enumerate}
\usepackage{enumitem}
\usepackage{float} 
\usepackage{mwe}
\usepackage{epstopdf}
\usepackage{diagbox}
\usepackage{fix-cm}
\usepackage{subcaption}
\usepackage{booktabs}
\usepackage{makecell}
\usepackage{extpfeil}
\usepackage{array}  
\usepackage{caption}
\usepackage{rotating}
\usepackage[round]{natbib}
\usepackage{authblk}
\usepackage{xcolor}
\definecolor{revGreen}{RGB}{0,130,0}
\definecolor{revYellow}{RGB}{200,140,0}
\definecolor{revRed}{RGB}{200,0,0}

\definecolor{revOrange}{RGB}{217,119,6}

\theoremstyle{plain}
\newtheorem{theorem}{Theorem}[section]

\newtheorem{lemma}[theorem]{Lemma}
\newtheorem{proposition}[theorem]{Proposition}
\theoremstyle{definition}

\theoremstyle{remark}
\newtheorem{remark}{Remark}
\theoremstyle{definition}
\newtheorem{assumption}{Assumption}

\newenvironment{keywords}
  {\small\noindent\textbf{Keywords:}}
  {\par\vspace{6pt}}
\newenvironment{classcode}
  {\small\noindent\textbf{JEL Classification:}}
  {\par}

\begin{document}

\title{Understanding Short-term Implied Volatility Dynamics: A Distribution-based Approach Beyond Stochastic Volatility}

\author[1,2]{Liexin Cheng}
\author[1,2]{Xue Cheng\thanks{Corresponding author. Email: chengxue@pku.edu.cn}}
\affil[1]{School of Mathematical Science, Peking University, China}
\affil[2]{Center for Statistical Science \& LMEQF, Peking University, China}
\date{}

\maketitle

\begin{abstract}
This paper examines the short-term asymptotic behavior of the implied volatility surface, focusing on the at-the-money (ATM) skew and curvature. Rather than committing to a specific stochastic differential equation, we adopt a distribution-based approach by imposing cumulant conditions on the log-return distribution. Under these weak assumptions, we derive a quadratic expansion of implied volatility as a function of moneyness for near-the-money options and asymptotic expressions for ATM skew and curvature as time to maturity approaches zero, treating the decay rates of the third and fourth cumulants as independent parameters and introducing a marginal-type classifier. These results highlight differences in ATM asymptotic properties across different types of log return distributions and yield a unified, model-independent characterization of short-term smile dynamics in terms of the scaling laws of the marginal cumulants, covering regular/rough stochastic volatility and distribution-based ones like scalable gamma/CGMY martingale models alike from simple moment information. We subsequently present a distribution-based calibration method that not only effectively validates the analytical approximations, but also exhibits strong interpretability and consistent performance, and discuss potential connections to model-independent path-dependent applications via martingale optimal transport. Overall, our findings provide model-independent analytical tools for evaluating model performance against market stylized features, accurately approximating short-term option prices, and performing robust calibration.
\end{abstract}


\begin{classcode}G12; G13; C58\end{classcode}

\section{Introduction}
The Black-Scholes implied volatility (IV), derived from option prices, provides a model-free, scale-invariant measure of future asset volatility. This quantity, denoted by $\sigma_{\text{BS}}(k, \theta)$, where $k$ is log-moneyness and $\theta$ is time to maturity, is central to derivatives pricing and calibration. On the IV surface, the short-term near-the-money volatility points, with characteristics including at-the-money (ATM) skew and curvature, shape the geometry of the surface and account for the majority of trading liquidity. Their efficient calibration is essential for volatility modeling, pricing exotic derivatives, managing volatility risk, and informing trading strategies, particularly under rapid market regime shifts.

Empirical evidence has consistently revealed a power-law decay in the ATM skew (see \cite{bayer2016pricing, carr2003finite, gatheral2018volatility}): $$\partial_k \sigma_{\text{BS}}(0, \theta) := \left.\tfrac{\partial \sigma_{\text{BS}}(k,\theta)}{\partial k} \right|_{k = 0}\propto \theta^{H - 1/2}$$
with $H \in (0, \tfrac{1}{2})$ as $\theta \to 0.$ \footnote{The validity of these explosive asymptotics at extremely short maturities remains debated, see \cite{delemotte2023yet, guyon2023does}.} This market feature has spurred extensive research into the short-term asymptotic behavior of asset-price models. In particular, stochastic volatility models (SVMs), valued for their parsimonious replication of market features, exhibit divergent asymptotic behaviors: regular SVMs, which assume diffusive volatility dynamics, do not conform to the empirically observed power-law decay of implied volatility (see \cite{berestycki2004computing, alos2007short,  forde2012small, pagliarani2017exact, friz2018option, karami2018approximation}). Rough SVMs, by contrast, reproduce this power-law behavior (see \cite{fukasawa2017short, bayer2019short, euch2019short, forde2021small, friz2022short}), as do Lévy-type jump models (see \cite{figueroa2016short, gerhold2016small}). This asymptotic agreement has lent strong empirical support to the rough-volatility paradigm (\cite{gatheral2018volatility, fukasawa2021volatility}) and provides a theoretical foundation for the calibration procedures built upon it (see \cite{guyon2021smile, abi2025volatility, ait2021implied}).

Despite these advances in SVMs, research on short-term IV asymptotics is restricted in two important ways. First, the existing literature predominantly couples the IV-surface analysis to specific process-level mechanisms, such as stochastic, local-stochastic, or rough volatility dynamics defined through stochastic differential equations (SDE). This coupling conflates a mechanism with what is observable from option prices: the short-time IV surface is determined by the scaling laws of the marginal cumulants of the log return, and any process that produces the same scaling produces the same leading-order skew and curvature. To illustrate, consider SVMs (both rough and non-rough) where ATM characteristics are tied to the cumulants of the underlying asset's log return distribution (see \cite{euch2019short}):
$$\partial_k \sigma_{\text{BS}}(0, \theta) \propto \tilde{\kappa}_3(\theta), \quad \partial_{k}^2 \sigma_{\text{BS}}(0,\theta) \propto \tilde{\kappa}_4(\theta) - 2\tilde{\kappa}_3^2(\theta),$$
where $\tilde{\kappa}_3(\theta)$ and $\tilde{\kappa}_4(\theta)$ denote the skewness (third cumulant) and excess kurtosis (fourth cumulant), respectively. Even when assuming higher-order cumulants ($\tilde \kappa_n$ for $n \geq 5$) are negligible, the asymptotic results are not universally valid. This observation suggests that imposing asymptotic conditions directly on the cumulants of the log-return distribution provides a more primitive foundation for short-time IV analysis than relying on a specific process-level SVM specification. 

Second, existing literature on SVM implied volatility asymptotics inadequately addresses the richer behavior of ATM curvature $\left(\partial_k^2 \sigma_{\text{BS}}(0, \theta)\right)$. Studies in \cite{alos2017curvature} and \cite{euch2019short} demonstrate that SVMs (particularly rough SVMs with Hurst parameter $H$) exhibit rigid asymptotic scaling: both ATM skew ($\partial_k \sigma_{\text{BS}}(0,\theta) \propto \theta^{H - 1/2}$) and curvature ($\partial_k^2 \sigma_{\text{BS}}(0,\theta) \propto \theta^{2H - 1}$) follow fixed power-law relationships. While this captures certain market behaviors, the rigid 2:1 ratio between curvature and skew explosion rates fails to match the dynamics observed in real option markets. Capturing this richer behavior requires decoupling the cumulant scaling rates of skewness and kurtosis rather than tying them to a single Hurst index.

To address these limitations, we propose a distribution-based framework that imposes asymptotic conditions directly on the cumulants of log returns. Our work shares with \cite{euch2019short,fukasawa2017short} the view that the cumulant content of the log return drives the leading-order behavior of the short-time implied-volatility surface: \cite{euch2019short} makes this explicit through a small-time Edgeworth expansion of the log-return density, while \cite{fukasawa2017short} extracts the same content in the form of an ATM-skew formula whose magnitude is governed by the log-return's third cumulant. Both works are built on, and confined to, stochastic volatility models specified by SDEs, which ties the cumulant scaling rates to a Hurst index and forces the $2{:}1$ curvature-to-skew explosion rate. Our framework fundamentally departs by introducing a distribution-based approach that frees the cumulant decay rates of skewness and kurtosis and introduces a marginal-type classifier $(m,n)$ to the log-return distribution. This also allows us to uncover novel and universal implied-volatility regimes inaccessible to the existing literature. Specifically,
\begin{enumerate}[itemsep=1pt,parsep=2pt,topsep=2pt]
    \item Arbitrary decay rates for skewness and kurtosis: by treating these decay rates as independent parameters of the framework, we uncover short-time IV regimes inaccessible to existing approaches;
    \item Generalized distributional assumptions: our framework is concerned with log-return marginal distributions parametrized by $(m,n)$, which classify the marginal type beyond the specific forms required by existing SVM-based approaches;
    \item Path independence: we eliminate the need for pathwise volatility dynamics and establish our main results solely on distributions.
\end{enumerate}  
In the following, we elaborate on these features and present our main results.

First, we generalize ATM asymptotics by allowing arbitrary, independent decay rates of skewness ($\beta_1 > 0$) and kurtosis ($\beta_2 > 0$). The ATM skew takes the form:
$$\partial_k \sigma_{\text{BS}}(0,\theta) \propto \kappa_3(\theta)\,\theta^{\beta_1 - 1/2} + (2-n)\kappa_2(\theta)\kappa_4(\theta)\,\theta^{\beta_2},$$
where $\kappa_2(\theta)$, $\kappa_3(\theta)$, $\kappa_4(\theta)$ are bounded and $m, n \in \mathbb{R}$ are parameters classifying the type of the marginal distribution. The results are shown to apply to regular and rough SVMs as special cases, and to SVMs with decaying leverage effect as $\theta \to 0$, with validation via numerical examples. An empirical analysis of S\&P~500 index options further reveals curvature-to-skew explosion ratios exceeding $2{:}1$, confirming the practical relevance of this generalization.

Second, we extend beyond SVMs by relaxing distributional constraints for asset prices. We assume that the log return marginals admit low-order cumulant expansions classified by parameters $(m, n)$. Based on the type of the marginal distributions, we derive asymptotic results on option price/implied volatility as well as ATM skew and curvature. We illustrate this generalization via two distribution-based models (scalable gamma/CGMY martingale models), where $\beta_1, \beta_2 > 0$ are freely specified and $(m, n) = (0, 0)$. By defining models purely through marginal distributions (hence ``distribution-based"), we leverage Kellerer’s theorem to ensure the existence of a semimartingale representation. Notably, when $\beta_1 - \tfrac{1}{2} > \beta_2$, the kurtosis term dominates ATM skew at a rate $\beta_2$, contrary to SVMs where $(m, n) = (1, 2)$ and the skewness term always dominates the ATM skew. Such dominance results are also illustrated in numerical experiments of the scalable gamma/CGMY martingale models under various scenarios of decay rates.

Third, our distribution-based approach eliminates the need to specify volatility dynamics by deriving short-term asymptotics directly from the log return density or characteristic function. The framework thus unifies analysis across diverse model classes while retaining the computational tractability of short-term IV asymptotics. Built on this feature, we propose a purely distribution-based calibration method that is exposed to little model risk. The calibration is performed on two-month implied volatility points, and effectively captures the market's features including ATM skew and curvature. Moreover, the calibrated parameters are interpreted as distribution characteristics and display consistency over the whole period.

In summary, this work addresses critical limitations in short-term implied volatility asymptotics through three contributions. First, it generalizes existing results by decoupling the decay rates of skewness ($\beta_1 > 0$) and kurtosis ($\beta_2 > 0$), accommodating ATM curvature behavior beyond the $2{:}1$ scaling of stochastic volatility models. Second, it introduces the $(m,n)$ classification of log-return marginals, unifying regular SVMs, rough SVMs, decaying-leverage SVMs, and distribution-based martingale models within a single framework, and showing that kurtosis can dominate ATM skew in certain model classes. Third, the distribution-based calibration requires no SDE specification and yields parameters with direct distributional meaning, providing a robust pricing and calibration tool with little exposure to model risk; it further constitutes a transparent first step toward model-independent hedging bounds via martingale optimal transport.

This article is structured as follows. In Section~\ref{sec 2}, we review the framework of \cite{euch2019short}, propose our distribution-based assumptions, and derive results on the IV expansion and ATM asymptotics. Sections~\ref{dist ver} and~\ref{svm ver} verify the framework on distribution-based martingale models and SVMs respectively. Section~\ref{calibration} presents the distribution-based calibration methodology. Other details of proofs are collected in the appendix.

\section{Distribution-based Asymptotics for the Implied Volatility Surface}
\label{sec 2}
In what follows, we first review the framework and results of \cite{euch2019short} to establish notation and motivation, then propose a distribution-based condition on the log-return marginals and derive a series of results: a uniform approximation of the option price, a near-the-money IV expansion, and ATM asymptotics of volatility skew and curvature.

Following the notation of \cite{euch2019short}, we consider the financial asset price process under a risk-neutral filtered probability space $(\Omega, \mathcal{F}, \{\mathcal{F}_t\}, Q)$ with $t\in [0,1]$:
$$S_t = S_0 e^{rt + Z_t},$$
where $r$ is the constant risk-free interest rate and $Z$ denotes the log return process of asset price $S$, satisfying $E[\exp(Z_t)] = 1$ for all $t \in [0,1]$. We fix current time $t = 0$ and let time to maturity be $\theta$. $F = S_0 e^{r\theta}$ denotes the forward price of $S$. Thus, the implied volatility of a European option with strike $K$ admits the representation as a function of log-moneyness and maturity: $\sigma_{\text{BS}} \equiv \sigma_{\text{BS}}(k, \theta)$ with $k = \log(K/F)$.

\subsection{Process-based Asymptotics under Stochastic Volatility of \cite{euch2019short}}
\label{sec:efgr-background}

Since our notation, key definitions, and motivating examples are built on \cite{euch2019short}, we begin by reviewing their setting and main results, then explain precisely where our framework extends beyond theirs.

In the context of SVMs of the form
\begin{equation}
\label{El model}
\mathrm{d}Z_t = -\tfrac{1}{2}v_t\,\mathrm{d}t + \sqrt{v_t}\,\mathrm{d}B_t,\quad v_t = v(Y_t),
\end{equation}
where $B$ is an $\mathcal{F}$-standard Brownian motion and $v$ is a positive continuous process adapted to a smaller filtration $\{\mathcal{G}_t\}_{t\ge 0}$, with $\mathrm{d}B_t = \rho_t\,\mathrm{d}W_t + \sqrt{1-\rho_t^2}\,\mathrm{d}W^\prime_t$, where $W^\prime$ is an $\mathcal{F}$-standard Brownian motion independent of $W$ and $(\rho_t)$ is a $\{\mathcal{G}_t\}$-progressively measurable process with $|\rho_t|\le1$. The following regularity condition is imposed:

\begin{assumption}
\label{assump:EFGR}
For any $p \ge 0$,
$$\sup_{\theta\in (0,1)}\left\|\tfrac{1}{\theta}\int_0^\theta v_t\,\mathrm{d}t\right\|_p < \infty, \qquad \sup_{\theta \in (0,1)} \left\| \left\{ \tfrac{1}{\theta} \int_0^\theta v_t (1 - \rho_t^2) \, \mathrm{d}t \right\}^{-1} \right\|_p < \infty,$$
where $\|\cdot\|_p$ is the $L^p$ norm under $Q$.
\end{assumption}

Throughout this paper $H_n$ denotes the $n$-th Hermite polynomial, defined by the recurrence $H_0(x)=1$, $H_1(x)=x$, $H_{n+1}(x)=xH_n(x)-nH_{n-1}(x)$; explicitly,
$$H_2(x)=x^2-1,\quad H_3(x)=x^3-3x,\quad H_4(x)=x^4-6x^2+3.$$

The proxy standard deviation is defined as
$\tilde{\sigma}_0(\theta) := \sqrt{\int_0^\theta E[v_t]\,\mathrm{d}t},$
abbreviated $\tilde\sigma_0$ when there is no confusion. An Edgeworth-adjusted density is then introduced, with bounded functions $\kappa_3(\theta)$ and $\kappa_4(\theta)$ encoding the skewness and excess kurtosis of $X_\theta$:
\begin{equation}
\label{q tilde}
\begin{aligned}
\tilde{q}_\theta(x) ={}& \phi\!\left(x + \tfrac{\tilde\sigma_0}{2}\right)\!\left\{1 + \kappa_3(\theta)\!\left(H_3\!\left(x + \tfrac{\tilde\sigma_0}{2}\right) - \tilde\sigma_0 H_2\!\left(x + \tfrac{\tilde\sigma_0}{2}\right)\right)\theta^H\right\} \\
& + \phi(x)\!\left(\kappa_4(\theta) H_4(x) + \tfrac{\kappa_3(\theta)^2}{2}H_6(x)\right)\theta^{2H},
\end{aligned}
\end{equation}
where $\phi$ is the standard normal density and $H \in (0,\tfrac{1}{2}]$. The put-price approximation functional is defined as:
\begin{equation}
\label{P tilde}
\begin{aligned}
\widetilde{P}(z,\theta) :={}& \tfrac{1}{\tilde\sigma_0}\!\left[\Phi\!\left(z+\tfrac{\tilde\sigma_0}{2}\right)e^{\tilde\sigma_0 z} - \Phi\!\left(z-\tfrac{\tilde\sigma_0}{2}\right)\right]\\
&+\kappa_3(\theta)\phi\!\left(z+\tfrac{\tilde\sigma_0}{2}\right)H_1\!\left(z+\tfrac{\tilde\sigma_0}{2}\right)e^{\tilde\sigma_0 z}\theta^H \\
&+\kappa_4(\theta)\phi\!\left(z+\tfrac{\tilde\sigma_0}{2}\right)\!\left[H_2\!\left(z+\tfrac{\tilde\sigma_0}{2}\right)+\tilde\sigma_0 H_1\!\left(z+\tfrac{\tilde\sigma_0}{2}\right)\right]e^{\tilde\sigma_0 z}\theta^{2H}
+\phi(z)\tfrac{\kappa_3(\theta)^2}{2}H_4(z)\theta^{2H}.
\end{aligned}
\end{equation}
The law of $X_\theta$ is further assumed to admit a density $p_\theta$ that satisfies the following approximation condition: for some $\alpha > \tfrac{5}{4}$,
\begin{equation}
\label{EFGR density cond}
\sup_{x\in\mathbb{R}}(1+x^2)^\alpha|p_\theta(x)-\tilde{q}_\theta(x)| = o(\theta^{2H}).
\end{equation}
Under Assumption \ref{assump:EFGR} and this condition, \cite{euch2019short} state that the following three results hold:

\label{thm:EFGR2}\noindent\textbf{(1)} For any $z_0\in\mathbb{R}$, the put option price $P(K, \theta)$ admits the approximation
\[
\frac{P(Fe^{\tilde\sigma_0 z},\theta)}{Fe^{-r\theta}\tilde\sigma_0} = \widetilde{P}(z,\theta)+o(\theta^{2H})
\]
uniformly for all $z\le z_0$, where $\widetilde{P}(z,\theta)$ is given in \eqref{P tilde}.

\label{thm:EFGR3}\noindent\textbf{(2)} For any $z\in\mathbb{R}$,
\begin{equation*}
\begin{aligned}
\sigma_{\mathrm{BS}}(\sqrt\theta\,z,\theta) = \tfrac{\tilde\sigma_0}{\sqrt\theta}\bigg\{1 &+\kappa_3(\theta)\!\left(\tfrac{z\sqrt\theta}{\tilde\sigma_0}+\tfrac{\tilde\sigma_0}{2}\right)\theta^H +\kappa_3(\theta)^2\!\left(\tfrac{3}{2}-\tfrac{3z^2\theta}{\tilde\sigma_0^2}\right)\theta^{2H}\\
&+\kappa_4(\theta)\!\left(-1+\tfrac{z^2\theta}{\tilde\sigma_0^2}\right)\theta^{2H}\bigg\}+o(\theta^{2H}).
\end{aligned}
\end{equation*}

\label{thm:EFGR4}\noindent\textbf{(3)} The ATM skew and curvature are defined as the first- and second-order partial derivatives of implied volatility at $k = 0$. They admit the approximations:
\begin{equation}
\label{EFGR asymptotics}
\partial_k\sigma_{\mathrm{BS}}(0,\theta) = \kappa_3(\theta)\,\theta^{H-1/2}+o(\theta^{2H-1/2}),\qquad
\partial_k^2\sigma_{\mathrm{BS}}(0,\theta) = 2\tfrac{\kappa_4(\theta) - 3\kappa_3(\theta)^2}{\tilde\sigma_0/\sqrt\theta}\,\theta^{2H-1}+o(\theta^{2H-1}).
\end{equation}

The condition~\eqref{EFGR density cond} can be verified for regular and rough SVMs. In particular, for the regular diffusion SVM where $v_t=v(Y_t)$ and $Y$ follows
\begin{equation}
\label{SVM regular}
\mathrm{d}Y_t = b(Y_t)\,\mathrm{d}t + c(Y_t)\,\mathrm{d}W_t,\quad \mathrm{d}B_t\,\mathrm{d}W_t=\rho\,\mathrm{d}t,\quad \rho\in(-1,1),
\end{equation}
with mild regularity on functions $v,b,c$, Assumption~\ref{assump:EFGR} and Condition~\eqref{EFGR density cond} hold with $H=\tfrac{1}{2}$. The cumulant functions are explicitly:
\begin{equation*}
\kappa_3(\theta) = \tfrac{\rho g(Y_0)}{2f(Y_0)},\qquad \kappa_4(\theta) = \tfrac{(1+2\rho^2)}{6}\tfrac{g^2}{f^2}(Y_0),
\end{equation*}
where $f=\sqrt{v}$ and $g=f'c$. Consequently Results~(1)--(3) apply, and by \textnormal{Eq.}~\eqref{EFGR asymptotics} both ATM skew and curvature converge to constants as $\theta\to 0$ (asymptotic order 0). For rough volatility models with $H\in(0,\tfrac{1}{2}]$, the cumulants scale as $\theta^H$ and $\theta^{2H}$, and the ATM skew and curvature explode at rates $H-\tfrac{1}{2}$ and $2H-1$ respectively.

\subsection{Distribution-based Framework: Cumulant Conditions and ATM Expansions}
\label{sec:main-framework}

The results above establish a fundamental insight: short-time ATM skew and curvature are governed by the low-order cumulants $\kappa_3(\theta)$ and $\kappa_4(\theta)$ and their decay rates, with the cumulant structure rather than the specific volatility dynamics determining the short-time behavior. However, two structural constraints embedded in this framework limit its scope.

First, the cumulant scaling rates are constrained to $(H,2H)$, forcing the curvature-to-skew explosion rate to be exactly $2{:}1$. This rigidity is a feature of the assumed pathwise mechanism rather than of the IV surface itself: imposing the scaling rates $(\beta_1,\beta_2)$ directly on the marginal cumulants is the more primitive assumption, since the short-time ATM asymptotics follow from it without depending on any process-level specification. Empirically, the S\&P~500 curvature-to-skew explosion ratio substantially exceeds $2{:}1$, a regime no fixed-$H$ rough SVM can reach. We accordingly free the cumulant scaling rates and treat $(\beta_1,\beta_2)$ as independent parameters of the framework, recovering the rough-SVM constraint as the special case $\beta_2=2\beta_1=2H$.

Second, there is lack of freedom in the Edgeworth series assumption of $\tilde q_{\theta}$. Although the approximating density $\tilde{q}_\theta$ is not a pure Edgeworth series: it contains an adjustment term $-\tilde\sigma_0 H_2(\cdot)\theta^H$ in the skewness correction, for the usual SVM setup with $H$ and $2H$ scaling these adjustments are asymptotically negligible. Once the cumulant scaling rates are freed, however, their form can affect leading-order ATM asymptotics and must be chosen with care. We accordingly generalize the framework by introducing free correction parameters $m,n\in\mathbb{R}$.

These relaxations together produce a distribution-based framework: rather than specifying a driving SDE, the framework's modeling assumptions are purely distributional, consisting of an Edgeworth-type approximating density family $q_\theta^{(m,n)}$ together with a density-convergence condition on log-return marginals. This setup can be shown to cover regular and rough SVMs, certain distribution-based martingale models, and jump models.

\begin{assumption}
    \label{assump:1} For $\theta \in [0, 1]$, the standard deviation of $Z_\theta$, denoted by $\sigma_0(\theta)$,  exists and satisfies $\sigma_0(\theta) = O(\sqrt{\theta})$ as $\theta \to 0$.
\end{assumption}

Let $X_\theta := \tfrac{Z_\theta}{\sigma_0(\theta)}$ (with $X_0 = 0$). For brevity, we let $\sigma_0 \equiv \sigma_0(\theta)$, $\kappa_3 \equiv \kappa_3(\theta)$, $\kappa_4 \equiv \kappa_4(\theta)$, and $\kappa_2 \equiv \tfrac{\sigma_0(\theta)}{\sqrt{\theta}}$. Then the price of a put option $P(K, \theta)$ satisfies, for every $z \in \mathbb{R}$,
\begin{equation}
\label{put price}
\tfrac{P(Fe^{z\sigma_{0}(\theta)},\theta)}{F\sigma_{0}(\theta)} = e^{-r\theta} \int_{-\infty}^{z} Q(\zeta \geq X_{\theta}) e^{\sigma_{0}(\theta)\zeta} \, \mathrm{d}\zeta.
\end{equation}
We introduce the following two-parameter family that will be used to approximate the density $p_\theta(x)$ of $X_\theta$:
\begin{equation}
\label{q}
    \begin{aligned}
q_\theta^{(m,n)}(x) &= \phi\left(x + \tfrac{\sigma_0}{2}\right) \left\{1 + \kappa_3 \left(H_3(x + \tfrac{\sigma_0}{2}) - m\sigma_0(\theta) H_2(x + \tfrac{\sigma_0}{2})\right) \theta^{\beta_1}\right\}\\&
\quad \quad  + \phi\left(x+\tfrac{\sigma_0}{2}\right) \kappa_4 \left(H_4(x + \tfrac{\sigma_0}{2}) -n\sigma_0 H_3(x+\tfrac{\sigma_0}{2})\right)\theta^{\beta_2}\\
& \quad \quad + \phi(x)\tfrac{\kappa_3^2}{2}H_6(x) \theta^{2\beta_1},
\end{aligned}
\end{equation}
where $\beta_1, \beta_2 > 0$, $\kappa_3, \kappa_4$ are bounded functions of $\theta$, and $m, n \in \mathbb{R}$. When $m = n = 0$, $q^{(0,0)}_\theta(x)$ is a truncated Edgeworth series approximation of $p_\theta(x)$ with the third cumulant $\kappa_3(\theta) \theta^{\beta_1}$ and the fourth cumulant $\kappa_4(\theta) \theta^{\beta_2}$; in general, $q^{(m,n)}_\theta$ can be viewed as a truncated Edgeworth series with high-order adjustments.

The specific form of the adjustment terms in~\eqref{q} ($-m\sigma_0 H_2$ for the skewness and $-n\sigma_0 H_3$ for the kurtosis) is determined by how the put-price functional acts on the approximating density. Motivated by the put-price integral representation~\eqref{put price}, we introduce the pricing operator
\begin{equation}
\label{pricing op}
\mathrm{I}^2 f(z) = \int_{-\infty}^z\!\!\int_{-\infty}^\zeta f(x)\,\mathrm{d}x\; e^{\sigma_0\zeta}\,\mathrm{d}\zeta.
\end{equation}
Computing $\mathrm{I}^2$ of the skewness correction term (by integration by parts) yields
\begin{align*}
\mathrm{I}^2\!\left(\phi\!\left(x+\tfrac{\sigma_0}{2}\right)\kappa_3(H_3-m\sigma_0H_2)\theta^{\beta_1}\right)
&= \phi\!\left(z+\tfrac{\sigma_0}{2}\right)\kappa_3\!\left(H_1\!\left(z+\tfrac{\sigma_0}{2}\right)+(1-m)\sigma_0\right)e^{\sigma_0z}\theta^{\beta_1}\\
&\quad+(m-1)\sigma_0^2\,\Phi\!\left(z-\tfrac{\sigma_0}{2}\right)\theta^{\beta_1}.
\end{align*}
An extra $\Phi(z-\sigma_0/2)$-type term $(m-1)\sigma_0^2\Phi(z-\sigma_0/2)\theta^{\beta_1}$ appears naturally under the operator $\mathrm{I}^2$, and an analogous term arises from the kurtosis correction. This justifies the specific form of $q_\theta^{(m,n)}$ given in~\eqref{q}.

Beyond this formal origin, the parameters $(m,n)$ play well-defined structural roles in $q_\theta^{(m,n)}$: $m$ scales the Hermite shift $-\sigma_0 H_2$ within the skewness term of order $\theta^{\beta_1}$, and $n$ scales the Hermite shift $-\sigma_0 H_3$ within the kurtosis term of order $\theta^{\beta_2}$. Probabilistically, these adjustments allow $q_\theta^{(m,n)}$ to capture finer structural features of the risk-neutral distribution beyond what the truncated Edgeworth series with cumulants $(\kappa_3, \kappa_4)$ alone can express. In financial terms, the skewness and kurtosis terms carry volatility-risk and vol-of-vol-type content respectively, and the parameters $(m,n)$ shape how these higher-order risk features enter the log-return density.

These structural roles propagate to the implied-volatility surface: the expansion in Theorem~\ref{Theorem 2} reveals that $m$ enters the skewness term at order ${\beta_1 + \tfrac{1}{2}}$ and $n$ enters the kurtosis term at order ${\beta_2 + \tfrac{1}{2}}$, so the two parameters shape the IV surface through their own density terms. Taking ATM derivatives carries these effects across in Theorem~\ref{ATM skew}: depending on the relative magnitudes of $(\beta_1,\beta_2)$, the $n$-bearing term can dominate the ATM skew and the $m$-bearing term can dominate the ATM curvature. Different $(m,n)$ therefore yield distinct implied-volatility shapes and identify different model classes, as established in Sections~\ref{dist ver} and~\ref{svm ver}.

Building on the approximating density $q_\theta^{(m,n)}$ defined in~\eqref{q}, we define the candidate put-price functional
$$
\begin{aligned}P^{(m,n)}(z, \theta) :=& \tfrac{1}{\sigma_0} \bigg[
    \Phi\left(z + \tfrac{\sigma_0}{2}\right) e^{\sigma_0 z} 
    - \left(1 + (1-m)\kappa_3\sigma_0^3\,\theta^{\beta_1}\right)\Phi\left(z - \tfrac{\sigma_0}{2}\right)
  \bigg] \\
& + \kappa_3 \phi\left(z + \tfrac{\sigma_0}{2}\right) 
    \left( H_1\left(z + \tfrac{\sigma_0}{2}\right) + (1 - m)\sigma_0 \right) 
    e^{\sigma_0 z} \theta^{\beta_1}  \\
& + \kappa_4\phi\left(z + \tfrac{\sigma_0}{2}\right) 
    \left( H_2\left(z + \tfrac{\sigma_0}{2}\right) + (1 - n)\sigma_0 H_1\left(z+\tfrac{\sigma_0}{2}\right) \right)e^{\sigma_0z}\theta^{\beta_2}  \\
&
    + \phi(z)\tfrac{\kappa_3^2}{2} H_4(z)\,\theta^{2\beta_1},
\end{aligned} $$ 
which extends the put-price functional~\eqref{P tilde} of \cite{euch2019short} to general $(m,n,\beta_1,\beta_2)$. Based on $P^{(m, n)}(z, \theta)$, we have an asymptotic approximation of the put option price if the density of $X_\theta$ is well-approximated by $q^{(m,n)}_\theta$:
\begin{theorem}
    \label{Th1}
    Under Assumption \ref{assump:1}, we assume that $X_\theta$ admits a density function $p_\theta(x)$ for every $\theta \in (0,1)$ and that there exists some $\alpha > 1$ and $m, n\in \mathbb{R}$ with
        \begin{equation}
        \label{density convergence}
\sup_{x \le x_0} (1 + x^2)^{\alpha} |p_\theta(x) - q^{(m,n)}_\theta(x)| = o(\theta^{\bar{\beta}_\epsilon}), \quad \text{as } \theta \to 0,
\end{equation}
for every $x_0 \in \mathbb{R}$ and $\epsilon > 0$, where 
\begin{equation}
    \label{eq::beta exp}
    \bar{\beta}_\epsilon = \min\{\beta_2 + \tfrac{1}{2}, 2\beta_1, 2\beta_2-\epsilon, \beta_1 + \beta_2 - \epsilon\}.
\end{equation}
Then, for any $z_0 \in \mathbb{R}$, the equality
\[\begin{aligned}
&\tfrac{p(F e^{\sigma_0 z}, \theta)}{F e^{-r\theta} \sigma_0}= P^{(m,n)}(z,\theta) + o(\theta^{\bar{\beta}_\epsilon})
\end{aligned}
\]
holds uniformly for all \(z \leq z_0\).
\end{theorem}
The proof of Theorem \ref{Th1} is given in Appendix \ref{proof Th1}.
\begin{remark}
\label{rem:Th1}
Theorem~\ref{Th1} extends the density-approximation result of \cite{euch2019short} along three lines. First, the freedom of $(m,n)$ enables different model classes to satisfy condition~\eqref{density convergence}: as will be shown later, $(m,n)=(0,0)$ corresponds to the distribution-based models of Section~\ref{dist ver}, while $(m,n)=(1,0)$ and $(1,2)$ correspond to SVMs without and with decaying leverage, respectively. This is also the source of the kurtosis effect on ATM skew revealed in Theorem~\ref{ATM skew}, an effect that is obscured under the fixed Hermite-coefficient form there. Second, the freedom of $(\beta_1,\beta_2)$ allows the framework to analyze a broader range of log-return distributions, including SVMs with decaying leverage as $\theta\to0$, and jump models whose log return has mild skewness so that kurtosis can dominate the ATM skew. Third, the density-convergence condition~\eqref{density convergence} is itself technically weakened: the supremum is taken over a left half-line rather than all of $\mathbb{R}$, and the polynomial weight is relaxed from $\alpha > \tfrac{5}{4}$ to $\alpha>1$.
\end{remark}

We note one limitation of the framework: condition~\eqref{density convergence} is Edgeworth-like, requiring the density to be well-approximated by $q_\theta^{(m,n)}$ up to four leading cumulants. General exponential Lévy models with high jump activity may violate this because the jump activity index can drive higher-order cumulants at incompatible rates.

Next, we consider the IV expansion and ATM asymptotics when \textnormal{Eq.} \eqref{density convergence} is satisfied.

\begin{theorem}
\label{Theorem 2}
        Suppose \textnormal{Eq.} \eqref{density convergence} holds for some \(\alpha > 1\) with \(q_\theta^{(m,n)}\) of the form in \textnormal{Eq.} \eqref{q}. Then, for any \(z \in \mathbb{R}\) and $\epsilon > 0$,
\begin{equation}
\label{imp vol}
\begin{aligned}
\sigma_{\operatorname{BS}}(\sqrt{\theta} z, \theta) &= \kappa_2 \bigg\{ 1 + \kappa_3 \left( \tfrac{z}{\kappa_2} + \left(\tfrac{3}{2}-m\right)\kappa_2 \sqrt{\theta} +(m-1) \kappa_2^2\tfrac{\Phi(\tfrac{z}{\kappa_2}-\tfrac{\kappa_2\sqrt{\theta}}{2})}{\phi(\tfrac{z}{\kappa_2}-\tfrac{\kappa_2\sqrt{\theta}}{2})}\theta\right) \theta^{\beta_1}\\
&  + \kappa_3^2\left( \tfrac{3 }{2} - 3  \tfrac{z^2}{\kappa_2^2} \right) \theta^{2\beta_1}  + \kappa_4\left(-1 + \tfrac{z^2}{\kappa_2^2}+ (2-n)z\sqrt{\theta}\right)\theta^{\beta_2} \bigg\}+ o(\theta^{\bar{\beta}_\epsilon}).
\end{aligned}
\end{equation}
    \end{theorem}
The proof of Theorem \ref{Theorem 2} is given in Appendix \ref{proof Th2}.

\begin{theorem}
  \label{ATM skew}
       Suppose we have \textnormal{Eq.} \eqref{density convergence} for some \(\alpha > 1\) with \(q_\theta^{(m,n)}\) of the form \textnormal{Eq.} \eqref{q}. Then, for any $\epsilon > 0$,
\[
\partial_{k}\sigma_{\mathrm{BS}}(0,\theta) = \kappa_{3}\theta^{\beta_1-1/2}+ (2-n)\kappa_2\kappa_4\theta^{\beta_2}+ o(\theta^{\bar \beta^-_\epsilon}),
\]
where $\bar \beta^-_\epsilon = \min\{\beta_1 + \tfrac{1}{2} -\epsilon, \bar{\beta}_\epsilon - \tfrac{1}{2}\}$, and
\[
\begin{aligned}
&\partial_{k}^{2}\sigma_{\mathrm{BS}}(0,\theta) = \tfrac{2\kappa_4}{\kappa_2}\theta^{\beta_2 - 1} -\tfrac{6\kappa_3^2}{\kappa_2}\theta^{2\beta_1 - 1} + \kappa_2\kappa_3 \left(\tfrac{\sqrt{2\pi}}{2}(m-1) + \tfrac{1-2m}{8}\kappa_2 \sqrt{\theta}\right)\theta^{\beta_1} + o(\theta^{\bar{\beta}_\epsilon -1}).
\end{aligned}
\]
\end{theorem}
The proof of Theorem \ref{ATM skew} is given in Appendix \ref{proof Th3}.
\begin{remark}
\label{rem:comparison}
The asymptotic expansions for ATM skew and curvature in Theorem~\ref{ATM skew} incorporate additional terms absent in prior studies (e.g., \cite{alos2017curvature, alos2007short, euch2019short, fukasawa2017short, guyon2021smile}). The additional terms reveal that both skewness and kurtosis of the log return can lead the ATM skew or curvature, depending on the relationship between the orders of the cumulants. These novel results arise from the refined cumulant terms and relaxed scaling assumptions and showcase distinct asymptotic behaviors under different model setups.

Specifically, for ATM skew, the additional term $(2-n)\kappa_2\kappa_4\theta^{\beta_2}$ can dominate the leading-order asymptotics when $\beta_1 > \beta_2+\tfrac{1}{2}$ and $n\neq 2$. This term reveals a dominance effect of the log return's kurtosis rather than skewness on ATM skew, contrasting with existing results where such cross-effects were obscured.

For ATM curvature, the additional term $\tfrac{\sqrt{2\pi}}{2}(m-1)\kappa_2\kappa_3\theta^{\beta_1}$ serves as the leading-order term under conditions $1<\beta_1<\beta_2-1$ and $m\neq 1$. This also reveals a more involved dependence structure of ATM curvature on skewness and kurtosis, as opposed to existing results where kurtosis plays the leading role.
\end{remark}


\section{Verification on Distribution-based Models}
\label{dist ver}
In this section, we propose two representative models within the $(m,n) = (0,0)$ family, where we apply a truncated Edgeworth series to density approximation. These models illustrate the framework’s flexibility in characterizing the asymptotics of distribution-based modeling.

The distribution-based models discussed here are designed to replicate the marginal distributions of Lévy-type log returns while avoiding the explosive behavior of short-term implied volatility asymptotics. As shown in \cite{figueroa2016short}, for pure-jump exponential Lévy models, the rate of such explosions depends on the jump activity index  and cannot be captured solely by low-order cumulants of the log return. Consequently, traditional exponential Lévy models, whose distribution of $X_\theta$ at maturity belongs to the family of infinitely divisible distributions, often fail to satisfy condition \eqref{density convergence}. To incorporate such infinitely divisible marginals of Lévy-type modeling while preserving analytical tractability, we introduce maturity-dependent parameterization for the marginal distributions, which in turn ensures \textnormal{Eq.} \eqref{density convergence} by controlling the growth rates of high-order cumulants as maturity goes to zero.

By adopting the parameterization, the proposed models preserve and, in some cases, amplify the key advantages of exponential Lévy models in option pricing, such as mathematical tractability, flexible short-term ATM asymptotics, and heavy-tailed return distributions, albeit at the cost of losing the original path-dependent information. Importantly, Kellerer's theorem (see \cite{kellerer1972markov}) guarantees that, under appropriate parameter constraints, there exists a Markov martingale whose marginals match those of the distribution-based model. This ensures the proposed models naturally satisfy the martingale condition required for discounted asset prices.

Before proceeding, we propose the following stronger condition on characteristic functions, which is more verifiable in certain cases than the assumption on density function (\textnormal{Eq.} \eqref{density convergence}).
\begin{lemma}
\label{CHF condition}
    If the following condition is satisfied for the characteristic functions of $\{X_\theta\}_{\theta \in (0,1)}$: there exists $v > 0$ such that 
    $\sup_{\theta \in (0,1)}\, E[e^{-v X_\theta}] < \infty$ and, for every $\epsilon > 0$, \begin{equation}
        \label{cond 2}\int_{-\infty}^\infty\left| \varphi_X(u + \mathrm{i}v;\theta) - \varphi^{(m,n)}(u+ \mathrm{i}v;\theta)\right|\mathrm{d}u = o(\theta^{\bar{\beta}_\epsilon}),\end{equation}
    where $\bar{\beta}_\epsilon$ is given by \textnormal{Eq.} \eqref{eq::beta exp} and $$\begin{aligned}\varphi^{(m,n)}(u;\theta)& = \exp\left(-\tfrac{\mathrm{i}u\sigma_0}{2} - \tfrac{u^2}{2}\right)\bigg\{1 - \kappa_3\left(\mathrm{i}u^3 - mu^2\sigma_0\right) \theta^{\beta_1}\\
    &\quad \quad-\tfrac{\kappa_3^2}{2}u^6 \theta^{2\beta_1} + \kappa_4 \left(u^4 + \mathrm{i}u^3 n \sigma_0\right) \theta^{\beta_2}\bigg\},
    \end{aligned}$$
    then \textnormal{Eq.} \eqref{density convergence} holds and the conclusions of Theorem \ref{Th1} apply.
\end{lemma}
The proof of Lemma \ref{CHF condition} is given in Appendix \ref{proof of prop1}. 
\subsection{Scalable Gamma Martingale Model}
Consider a distribution-based model with returns the difference of two gamma distributions: for each $\theta \in (0,1]$, let
\begin{equation}
\label{gamma model}
     Z_\theta = Z^{(1)}_\theta - Z^{(2)}_\theta + w_\theta, 
\end{equation}
where $Z^{(1)}_\theta \sim \Gamma(k_\theta, \gamma_\theta^{-1})$, $Z^{(2)}_\theta \sim \Gamma(k_\theta + \bar{k}_\theta, \gamma_\theta^{-1})$. Meanwhile, $w_\theta$ is a normalizing constant:
\[ w_\theta = -\ln E\left[ \exp\left(Z^{(1)}_\theta - Z^{(2)}_\theta \right) \right], \]
and, for the density $p_\theta^Z(x)$ of $e^{Z_\theta}$, we impose the following assumption \footnote{This assumption is numerically validated in Appendix \ref{cond Kellerer} for the experiments in Figure \ref{num_exam}.}
\begin{equation}
\label{convex order}\int f(x) p_s^Z(x)\mathrm{d}x \le \int f(x)p_t^Z(x) \mathrm{d}x, \quad 0< s \le t \le 1,
\end{equation}
for any convex function $f(\cdot)$, which ensures the existence of a Markov martingale whose marginals match those of $\{\exp{(Z_t)}\}_{t \in [0, 1]}$ by Kellerer's theorem.

A special case of this model is presented in \cite{geman2002pure}, where $\{Z_\theta^{(1)}\}_{\theta \in (0, 1]}$ and $\{Z_\theta^{(2)}\}_{\theta \in (0, 1]}$ correspond to the marginals of a gamma process. To derive a distinctive asymptotic structure, we further give some parameter scaling constraints:
the shape parameters $k_\theta = O(\theta^{\alpha})$, $\bar{k}_\theta = O(\theta^{\bar{\alpha}})$, and rate parameter $\gamma_\theta = O(\theta^{\beta})$ are set with $\alpha \in (-1, 0)$ $($to ensure decay rate $\beta_2 > 0$, as will be shown later$)$, $\bar{\alpha} > \alpha$ $($to ensure that $k_\theta$ dominates $\bar k_\theta$ as $\theta \to 0)$, and
$\beta = \tfrac{1}{2}(1 - \alpha)$ $($to achieve $\sigma_0(\theta) = O(\sqrt{\theta}))$.  
\begin{proposition}
\label{gamma}
Under the parameterization above and condition \eqref{convex order},  model \eqref{gamma model} satisfies the condition of Lemma \ref{CHF condition} with parameter $(m, n)= (0,0)$, $\beta_1 = \bar{\alpha} - \tfrac{3}{2}\alpha$ and $\beta_2 = -\alpha$. Furthermore, its Black-Scholes implied volatility exhibits the asymptotic behavior:
    \[
\partial_{k}\sigma_{\mathrm{BS}}(0,\theta) = \tfrac{\tilde{\kappa}_{3}}{6}\theta^{-1/2} + \tfrac{1}{12}\kappa_2\tilde{\kappa}_4 + o(\theta^{\bar{\beta}_\epsilon - 1/2})
\]
and
\[
\partial_{k}^{2}\sigma_{\mathrm{BS}}(0,\theta) = \tfrac{\tilde{\kappa}_4}{12\kappa_2}\theta^{-1} + o(\theta^{-\alpha -1}),
\]
where $\kappa_2 \equiv \kappa_2(\theta) = \tfrac{(2k_\theta + \bar{k}_\theta)^{1/2}\gamma_\theta}{\sqrt{\theta}}$ and 
$$
    \tilde{\kappa}_3 \equiv \tilde{\kappa}_3(\theta) = -\tfrac{2\bar{k}_\theta}{(2k_\theta + \bar{k}_\theta)^{3/2}},\quad 
    \tilde{\kappa}_4 \equiv \tilde{\kappa}_4(\theta) = \tfrac{6}{2k_\theta+ \bar{k}_\theta}
$$
are the skewness and excess kurtosis of $X_\theta$, respectively.
\end{proposition}
The proof is given in Appendix \ref{proof gamma}.

\subsection{Scalable CGMY Martingale Model}
Consider a distribution-based model with returns following \text{CGMY} marginal distributions: for each $\theta \in (0,1]$, let 
\begin{equation}
\label{cgmy model}
Z_\theta = Z_\theta^{(1)} + w_\theta,
\end{equation}
where $Z_\theta^{(1)} \sim \operatorname{CGMY}(C_\theta, G_\theta, M_\theta, Y)$. Meanwhile, $w_\theta$ is a normalizing constant:
\[ w_\theta = -\ln E\left[ \exp\left(Z^{(1)}_\theta\right) \right], \]
and, for the density $p_\theta^Z(x)$ of $e^{Z_\theta}$, we assume condition \eqref{convex order} to ensure the existence of a Markov martingale whose marginals match those of $\{\exp{(Z_t)}\}_{t \in [0, 1]}$ by Kellerer's theorem.

A special case of the model can be found in \cite{carr2002fine}, where $\{Z_\theta^{(1)}\}_{\theta \in (0, 1]}$ are the marginals of a \text{CGMY} process. To derive a distinctive asymptotic structure, we further give some parameter-scaling constraints:
the parameters $M_\theta = O(\theta^{\alpha_M})$, $G_\theta = M_\theta + m_\theta$ (with $m_\theta = O(\theta^{\alpha_G}) > 0$) and $C_\theta = O(\theta^{\beta})$ are set with $\alpha_M \in (-1, -0.5)$ $($to ensure decay rate $\beta_2 > 0$, as will be shown later$)$, $\alpha_G > \alpha_M$ $($to ensure that $M_\theta$ dominates $m_\theta)$, and $\beta =1- (Y-2)\alpha_M$ (to ensure $\sigma_0(\theta) = O(\sqrt{\theta}))$.

\begin{proposition}
\label{CGMY}
Under the parameterization above and condition \eqref{convex order}, model \eqref{cgmy model} satisfies the conditions of Lemma \ref{CHF condition} with $(m,n) = (0,0)$, $\beta_1 = \alpha_G - 2\alpha_M - \tfrac{1}{2}$ and $\beta_2 = -2\alpha_M - 1$. Furthermore, its Black-Scholes implied volatility exhibits the asymptotic behavior:
    \[
\partial_{k}\sigma_{\mathrm{BS}}(0,\theta) = \tfrac{\tilde{\kappa}_{3}}{6}\theta^{-1/2} + \tfrac{1}{12}\kappa_2\tilde{\kappa}_4 + o(\theta^{\bar{\beta}_\epsilon - 1/2}),
\]
and
\[
\partial_{k}^{2}\sigma_{\mathrm{BS}}(0,\theta) = \tfrac{\tilde{\kappa}_4}{12\kappa_2}\theta^{-1} + o(\theta^{-2\alpha_M - 2}),
\]
where 
$$\tilde{\kappa}_3 \equiv \tilde{\kappa}_3(\theta) = \tfrac{C_\theta \Gamma(3-Y)\left(M^{Y-3}_\theta - G^{Y-3}_\theta\right)}{\left(C_\theta\Gamma(2-Y)(M_\theta^{Y-2} + G_\theta^{Y-2})\right)^{\tfrac{3}{2}}},$$
$$\tilde{\kappa}_4 \equiv \tilde{\kappa}_4(\theta) = \tfrac{C_\theta \Gamma(4-Y)\left(M^{Y-4}_\theta + G^{Y-4}_\theta\right)}{\left(C_\theta\Gamma(2-Y)(M_\theta^{Y-2} + G_\theta^{Y-2})\right)^2}$$ are the skewness and excess kurtosis of $X_\theta$, respectively.
\end{proposition}
The proof is given in Appendix \ref{proof CGMY}.

\subsection{Numerical Experiments}
Model \eqref{gamma model} and \eqref{cgmy model} are constructed to allow for flexible short-term asymptotics in the implied volatility surface via the shape parameters of marginal distributions ($(\alpha, \bar{\alpha}, \beta)$ for model \eqref{gamma model} and $(\alpha_M, \alpha_G, \beta)$ for model \eqref{cgmy model}), which then results in the heterogeneous explosion rates of ATM skew and curvature.

\begin{figure}[!htbp]
    \centering
    \begin{subfigure}[b]{0.32\textwidth}
        \centering
        \includegraphics[width=\textwidth]{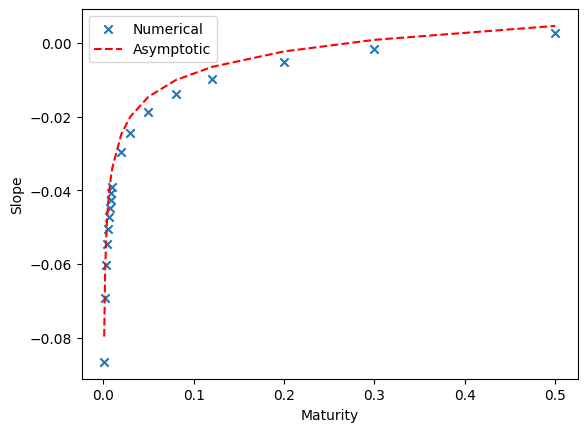} 
        \caption{$(\alpha, \bar \alpha)$= (-0.2, -0.1)}
    \end{subfigure}
    \hfill
    \begin{subfigure}[b]{0.32\textwidth}
        \centering
        \includegraphics[width=\textwidth]{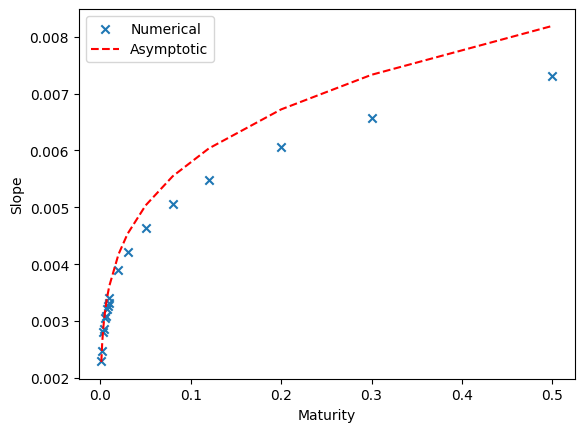} 
        \caption{$(\alpha, \bar \alpha)$ = (-0.2, 0.4)}
    \end{subfigure}
    \hfill
    \begin{subfigure}[b]{0.32\textwidth}
        \centering
        \includegraphics[width=\textwidth]{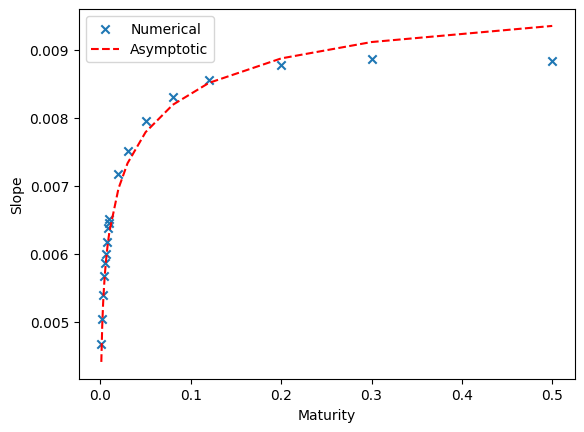} 
        \caption{$(\alpha, \bar \alpha)$ = (-0.2, 0.6)}
    \end{subfigure}
    
    \vspace{0.5cm} 
    
    \begin{subfigure}[b]{0.32\textwidth}
        \centering
        \includegraphics[width=\textwidth]{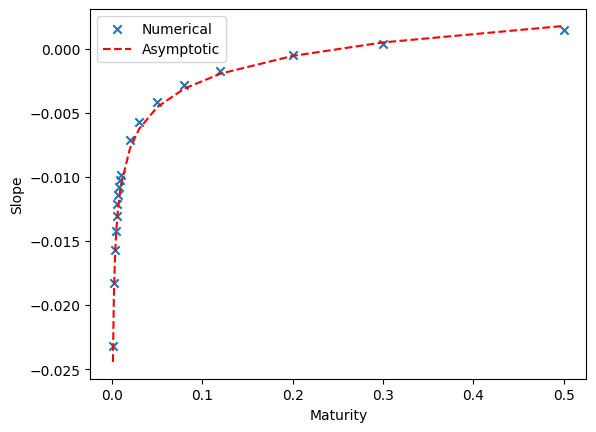}
        \caption{$(\alpha_M, \alpha_G)$ = (-0.6, -0.5)}
    \end{subfigure}
    \hfill
    \begin{subfigure}[b]{0.32\textwidth}
        \centering
        \includegraphics[width=\textwidth]{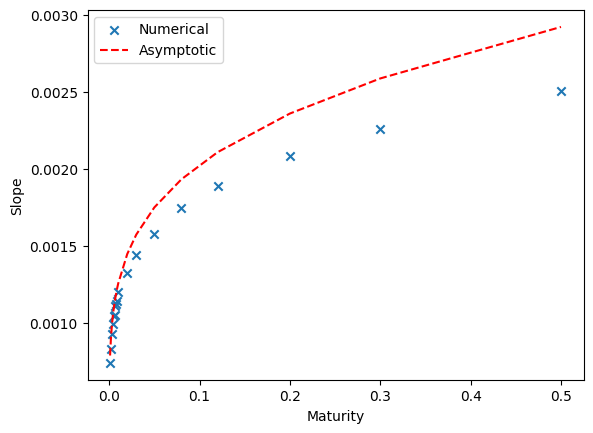} 
        \caption{$(\alpha_M, \alpha_G)$ = (-0.6, 0)}
    \end{subfigure}
    \hfill
    \begin{subfigure}[b]{0.32\textwidth}
        \centering
        \includegraphics[width=\textwidth]{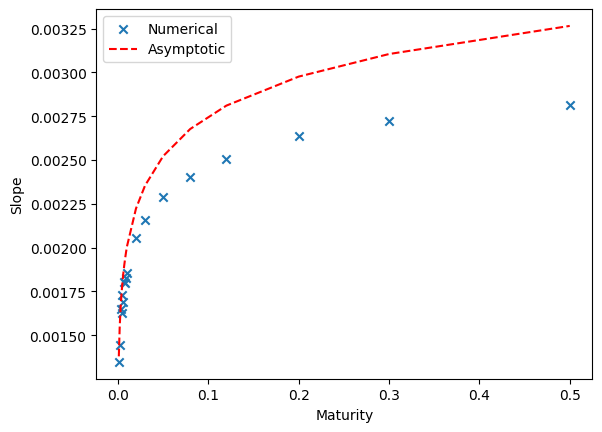}
        \caption{$(\alpha_M, \alpha_G)$ = (-0.6, 0.2)}
    \end{subfigure}
    
    \caption{Comparison of the theoretical asymptotics and the true values of ATM skew. The true values are computed as the numerical derivative of the ATM implied volatility, and the asymptotic values are computed according to Proposition \ref{gamma} (first row) and Proposition \ref{CGMY} (second row). The parameters of scalable gamma martingale model are set according to $(k_\theta, \bar{k}_\theta, \gamma_\theta) = (3\theta^{\alpha}, \tfrac{1}{2}\theta^{\bar \alpha}, \tfrac{1}{10}\theta^{\beta}).$ And the parameters of scalable CGMY martingale model are set according to $(M_\theta, G_\theta, C_\theta, Y) = (5\theta^{\alpha_M}, \tfrac{1}{2}\theta^{\alpha_G}, \tfrac{1}{10}\theta^{\beta}, 1.5).$ These constants are arbitrarily chosen, and other reasonable values yield similar patterns.}
    \label{num_exam}
\end{figure}

Figure \ref{num_exam} compares the theoretical asymptotics of ATM skew for the scalable Gamma/ CGMY martingale models with their corresponding true values. The validation of the increasing convex order condition, i.e., formula \eqref{convex order}, is shown in Figures \ref{gamma_kellerer} and \ref{cgmy_kellerer}, with validation details presented in Appendix \ref{cond Kellerer}.

The scenarios are categorized into three distinct cases, ordered from left to right, based on the dominant term in the asymptotics of ATM skew: $\tfrac{1}{6}\tilde{\kappa}_3 \theta^{-1/2}$ with the skewness term leading, $\tfrac{1}{6}\tilde{\kappa}_3 \theta^{-1/2} + \tfrac{1}{12}\kappa_2 \tilde{\kappa}_4$ with the combination of skewness and kurtosis terms leading, and $\tfrac{1}{12}\kappa_2 \tilde{\kappa}_4$ with the kurtosis term leading. A key observation from Figure \ref{num_exam} is that the $\tilde{\kappa}_4$ (excess kurtosis) term can drive the ATM skew to positive values despite the log return distribution actually having negative skewness. This highlights how higher-order moments, specifically kurtosis, can counteract the skewness effect, thereby shaping the short-term asymptotics of the volatility surface. The comparison underscores the distinct roles of skewness and kurtosis in determining the short-term implied volatility behavior.

\section{Verification and Extension on SVMs}
\label{svm ver}
In this section, we first demonstrate that the theoretical framework established in Theorem~\ref{Th1} encompasses the SVMs analyzed in \cite{euch2019short}. Specifically, we verify that these SVMs satisfy the density-convergence condition~\eqref{density convergence}, thus covering their results as a special case of ours. Building on this verification, we show that regular SVMs belong to the class $(m,n)=(1,0)$ and that SVMs with time-dependent leverage decay belong to the class $(m,n)=(1,2)$, a feature that helps explain the heterogeneous explosion rates of ATM skew and curvature in options markets.

\subsection{Verification on Conventional SVMs}
We consider the regular SVM given by \eqref{SVM regular}, with $\tilde{\sigma}_0(\theta)$ as defined in Section~\ref{sec:efgr-background}. In addition to Assumptions~\ref{assump:EFGR} and~\ref{assump:1}, the proof of Proposition~\ref{substitute} below requires the following variance condition.

\begin{assumption}
\label{assump:2}
$\displaystyle\lim_{\theta \downarrow 0} \operatorname{Var}\!\left(\tfrac{1}{\theta}\int_0^\theta v_t\,\mathrm{d}t\right) = 0$.
\end{assumption}

\begin{proposition}
\label{substitute}
     Under Assumptions~\ref{assump:EFGR}--\ref{assump:2}, with the log return governed by \textnormal{Eq.}~\eqref{SVM regular}, let $\tilde q_\theta$ be the Edgeworth-adjusted density defined in \textnormal{Eq.}~\eqref{q tilde}. Then for any $\alpha \in \mathbb{N} \cup \{0\}$, 
$$\sup_{x\in \mathbb{R}} (1 + x^2)^\alpha |q^{(1,0)}_\theta(x) - \tilde{q}_\theta(x)| = o(\theta^{2H}),$$
where $q_\theta^{(1,0)}(x)$ is given by~\eqref{q} with $\beta_1 = H$ and $\beta_2 = 2H$.
\end{proposition}
The proof is given in Appendix \ref{proof of prop2}. 

As established in Theorem~2.1 of \cite{euch2019short}, the SVMs considered in that work satisfy condition~\eqref{density convergence} when $q_\theta^{(m,n)}(x)$ is replaced by $\tilde q_\theta(x)$. Together with Assumption~\ref{assump:2}, Proposition~\ref{substitute} then implies that these SVMs are subsumed within our theoretical framework. In other words, our assumptions are less restrictive and encompass their modeling setup by setting $m=1$, $n=0$, $\beta_1=H$, and $\beta_2=2H$. This establishes a direct connection between their results and our generalized asymptotic analysis, and all our earlier conclusions, including Theorems~\ref{Th1}--\ref{ATM skew}, apply immediately to their setting; in particular, Results~(2)--(3) of Section~\ref{sec:efgr-background} are recovered and generalized to arbitrary $(\beta_1,\beta_2,m,n)$.

Specifically, our characterization of the asymptotic behavior of the implied volatility formalized in Theorems~\ref{Th1}--\ref{ATM skew} applies to the regular SVM~\eqref{SVM regular} introduced in Section~\ref{sec:efgr-background}. Moreover, certain rough volatility models such as rough Bergomi are also covered by Theorems~\ref{Th1}--\ref{ATM skew} as special cases, following from Section~4 of \cite{euch2019short}. This emphasizes the broad applicability of our analysis across both regular and rough SVM specifications.

\subsection{Extension to Decaying Leverage}
The rigid $2{:}1$ curvature-to-skew scaling $\theta^{H-1/2}:\theta^{2H-1}$ established in the rough-volatility literature~(\cite{alos2017curvature,euch2019short,forde2021small,friz2022short}) corresponds to the constraint $\beta_2=2\beta_1=2H$. This scaling does not match the patterns observed in real option markets, as we demonstrate using S\&P~500 index options.

\begin{figure}[!htbp]
    \centering
\includegraphics[width=0.6\linewidth]{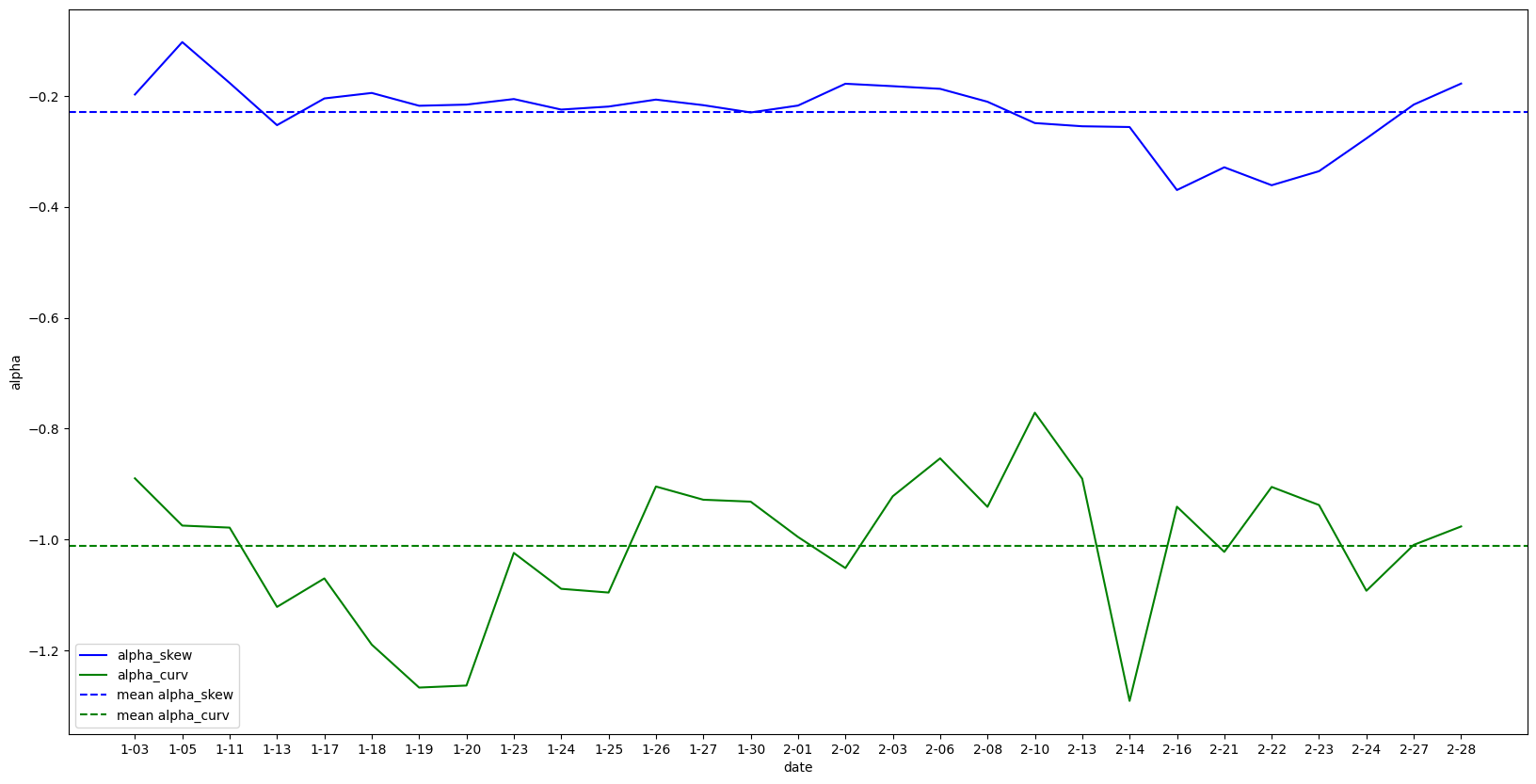}
\caption{The empirical ATM skew and curvature of S\&P 500 index options from 2023-01-03 to 2023-02-28. Details of the data processing and computation method can be found in Appendix \ref{data process}.}
\label{empirical}
\end{figure}

Figure~\ref{empirical} plots the empirical ATM skew and curvature of S\&P~500 index options from 2023-01-03 to 2023-02-28. For each expiry $\theta$, the forward price $F(\theta)$ is recovered from put--call parity using the call/put pair closest to the at-the-money strike $K$, $F(\theta) = K + e^{r(\theta)\theta}\bigl(C(K,\theta) - P(K,\theta)\bigr)$, with $r(\theta)$ the interpolated Treasury par yield at maturity~$\theta$. See Appendix~\ref{data process} for details of the data-filtering and computing methods. 

As shown in Figure~\ref{empirical}, the empirical ATM curvature of the S\&P~500 index market exhibits an explosion rate more than twice as fast as that of the ATM skew, a discrepancy unaccounted for by regular and rough SVMs. In fact, the curvature explosion rate can be nearly five times that of the skew, too significant to be attributed to computational errors.

To reconcile this empirical observation, we argue that the phenomenon suggests a time-dependent leverage decay mechanism that suppresses the explosion rate of the ATM skew in a given rough SVM. To formalize this intuition, we consider using $q_\theta^{(1,2)}(x)$ as the approximating candidate of the log return density. This choice, instead of $q^{(1,0)}_\theta(x)$, satisfies the accuracy condition in \textnormal{Eq.} \eqref{density convergence} and allows us to thoroughly analyze the ATM asymptotics under leverage decay. In the following, we exemplify this mechanism through SVMs with explicit leverage dynamics.

We consider the regular SVM~\eqref{SVM regular} with the decaying leverage $\rho_t = \rho\,t^{\alpha_\rho}$.
\begin{proposition}[Decaying Leverage]
\label{decaying leverage}
     Under Assumptions~\ref{assump:EFGR}--\ref{assump:2}, the SVM~\eqref{SVM regular} with leverage $\rho_t = \rho t^{\alpha_\rho}$ ($\rho\in(-1,1)$, $\alpha_\rho>0$) satisfies \textnormal{Eq.} \eqref{density convergence} with $(m,n) = (1,2)$, $\beta_1 = \tfrac{1}{2} + \alpha_\rho$, $\beta_2 =1$ and
    $$\kappa_3 \equiv \tfrac{\rho g(Y_0)}{2 f(Y_0)}, \quad \kappa_4 = {\tfrac{\rho_\theta^2}{6}}\tfrac{g^\prime c}{f}(Y_0) + {\tfrac{1 + 2\rho_\theta^2}{6}} \tfrac{g^2}{f^2}(Y_0),$$
where $f = \sqrt{v}$, $g = f^\prime c.$ Furthermore, the ATM skew and curvature take the values
    $$\partial_k \sigma_{\text{BS}}(0,\theta) = \kappa_3 \theta^{\alpha_\rho} + o(\theta^{\bar\beta - \tfrac{1}{2}})$$
    and 
    \[
\partial_{k}^{2}\sigma_{\mathrm{BS}}(0,\theta) = 2\tfrac{\kappa_4}{\kappa_2} + o(\theta^{\bar{\beta} - 1}),
\]
respectively, with $\bar{\beta} = \min\{1 + 2\alpha_\rho, 3/2\}$.
\end{proposition}

For the proof, please see Appendix \ref{proof decaying leverage}.

Proposition \ref{decaying leverage} shows that the skewness term dominates the ATM skew under the decaying leverage effect with $\alpha_\rho \in (0, 1]$. It also rules out the possibility that the kurtosis term $\kappa_2\kappa_4 \theta^{\beta_2}$ dominates the ATM skew asymptotics or that the skewness term dominates the ATM curvature.

\begin{figure}[!htbp]
    \centering
    \begin{subfigure}[b]{0.45\textwidth}
        \centering
        \includegraphics[width=\textwidth]{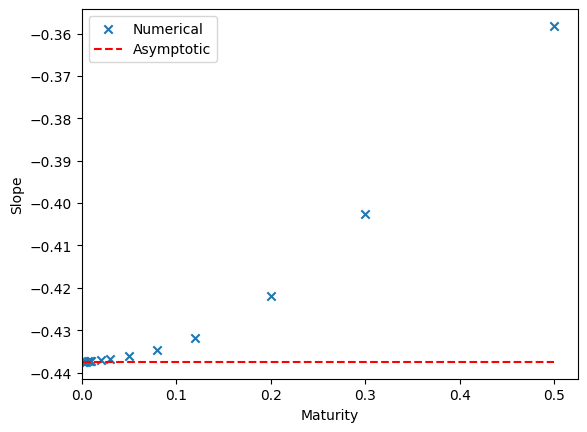} 
        \caption{$\alpha_\rho$= 0}
    \end{subfigure}
    \hfill
    \begin{subfigure}[b]{0.45\textwidth}
        \centering
        \includegraphics[width=\textwidth]{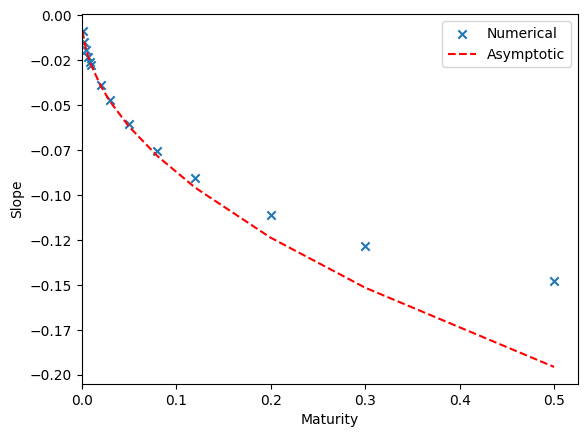} 
        \caption{$\alpha_\rho$ = 0.5}
    \end{subfigure}
    \caption{Comparison of the theoretical asymptotics and the true values of ATM skew in Heston model. The true values are computed as the numerical derivative of the ATM implied volatility, and the asymptotic values are computed according to Proposition \ref{decaying leverage}. The numerical derivative is calculated using near-the-money implied volatility points (i.e., $k \approx 0$). The Heston model takes parameters $(\kappa, \bar{v}, \eta, \rho, v_0) = (1, 0.06, 0.5, -0.7, 0.04).$ Other reasonable parameters also produce similar patterns.}
    \label{num_exam0}
\end{figure}

Figure \ref{num_exam0} compares the theoretical asymptotics and the true values (numerical derivative) of ATM skew in Heston model (i.e., $v_t \equiv Y_t$, $\mathrm{d}Y_t = \kappa (\bar{v} - Y_t)\mathrm{d}t + \eta \sqrt{Y_t}\mathrm{d}W_t$) with parameters $(\kappa, \bar{v}, \eta, \rho, v_0) = (1, 0.06, 0.5, -0.7, 0.04)$. The left subplot recovers the usual model setup with a constant leverage effect and the right subplot assumes a power-law decay leverage with $\alpha_\rho = 0.5$. 

The numerical results in Figure \ref{num_exam0} align with the leading effect of the skewness term as shown in Proposition \ref{decaying leverage} since the decay rate exactly matches the theoretical rate $\alpha_\rho$. Similar results with leverage decay can possibly be extended to certain rough volatility models, which have the practical strength of recovering the empirical asymptotics of ATM skew and curvature shown in Figure \ref{empirical}.

Table \ref{tab:models} further provides explicit asymptotics results for certain representative SVMs and distribution-based models discussed in this work. The columns from the left to the right are: model names, parameters in the approximating density $q^{(m,n)}_\theta$ including $(m, n)$, $\kappa_3(\theta)$ and $\kappa_4(\theta)$, and the asymptotic order of ATM skew and curvature. 

The table shows that regular SVMs, including Heston and 3/2 volatility models, have the short-term ATM skew and curvature converging to a constant. Rough Bergomi model has exploding rates $H- 1/2: 2H - 1$ for the ATM skew and curvature. For Heston model with leverage decay, an adjustment of parameter from $n = 0$ to $n = 2$ is needed to show the asymptotic order of $\alpha_\rho$ in the ATM skew. And a truncated Edgeworth series (i.e., $(m, n) = (0,0)$) is required for distribution-based models like scalable gamma/CGMY martingale model.

\captionsetup{hypcap=false}
\begin{center}
\renewcommand{\arraystretch}{1.2}

\rotatebox[origin=c]{90}{%
  \small
  \begin{minipage}{\textheight} 
    \centering
    \begin{tabular}{l c c c l}
    \toprule
    \textbf{Model} & $\boldsymbol{(m, n)}$  & $\boldsymbol{\kappa_3(\theta)}$ & $\boldsymbol{\kappa_4(\theta)}$ & \textbf{Asymptotic Order} \\
    \midrule
    Heston (cf. \cite{heston1993closed}) & $(1,0)$ & $\tfrac{\rho \eta}{4\sqrt{v_0}}$ & $\tfrac{(1 + 2\rho^2)\eta^2}{24v_0}$ & \makecell[c]{\text{Skew: 0}\\\text{Curv: 0} }\\
    3/2 (cf. \cite{carr2007new}) & $(1,0)$ & $\tfrac{\rho\eta\sqrt{v_0}}{4}$ & $\tfrac{(1 + 3\rho^2)\eta^2v^2_0}{6}$ & \makecell[c]{\text{Skew: 0}\\\text{Curv: 0} }\\
    $v_t = v(Y_t)$, $\mathrm{d}Y_t = b(Y_t)\mathrm{d}t + c(Y_t)\mathrm{d}W_t$ & $(1, 0)$ & $\tfrac{\rho}{2} \tfrac{g(Y_0)}{f(Y_0)}$ & $\tfrac{g^\prime c}{f}(Y_0)\tfrac{\rho^2}{6} + \tfrac{g^2}{f^2}(Y_0) \tfrac{1 + 2\rho^2}{6}$ &  \makecell[c]{\text{Skew: 0}\\\text{Curv: 0} }\\
    Rough Bergomi (cf. \cite{bergomi2005smile}) & $(1,0)$ & $\rho \eta \sqrt{\tfrac{H}{2}}\tfrac{I(\theta)}{\theta^{H+\tfrac{3}{2}} (E[v_\theta])^3}$ & $J(\theta)$ & \makecell[c]{\text{Skew: $H-\tfrac{1}{2}$}\\\text{Curv: $2H - 1$} }\\
    Heston with decaying leverage (Proposition \ref{decaying leverage}) & $(1,2)$ & $\tfrac{\rho \eta }{4\sqrt{v_0}}\theta^{\alpha_\rho}$ & {$\tfrac{(1 + 2\rho_\theta^2)\eta^2}{24v_0}$} & \makecell[c]{\text{Skew: $\alpha_\rho$}\\\text{Curv: 0} }\\
    Scalable gamma martingale (Proposition \ref{gamma}) & $(0,0)$ & $\tfrac{\tilde{\kappa}_3(\theta)}{6}\theta^{\tfrac{3}{2}\alpha - \bar \alpha}$ & $\tfrac{\tilde{\kappa}_4(\theta)}{24}\theta^\alpha$ & \makecell[c]{\text{Skew: $(\bar{\alpha} - \tfrac{3}{2}\alpha -\tfrac{1}{2})\wedge -\alpha$}\\\text{Curv: $-\alpha - 1$} }\\
    Scalable CGMY martingale (Proposition \ref{CGMY}) & $(0,0)$ & $\tfrac{\tilde{\kappa}_3(\theta)}{6}\theta^{2\alpha_M-\alpha_G+\tfrac{1}{2}}$ & $\tfrac{\tilde{\kappa}_4(\theta)}{24}\theta^{2\alpha_M + 1}$ & \makecell[c]{\text{Skew: $(- 2\alpha_M - 1) + \alpha_G \wedge 0$}\\\text{Curv: $-2\alpha_M - 2$} }\\
    \bottomrule
    \end{tabular}
    \vspace{1em}
    \captionof{table}[Short caption for list of tables]{A summary of short-term asymptotics for selected models. The column names $(m, n)$, $\kappa_3(\theta)$ and $\kappa_4(\theta)$ correspond to terms in $q^{(m,n)}_\theta(x)$ given by \textnormal{Eq.} \eqref{q}. In the table, the diffusive volatility model in the third line is from \cite{euch2019short} with $f=\sqrt{v}$ and $g = f^\prime c$; in the fourth line, the functions $I(\theta) = \int_0^\theta\int_0^t (t-s)^{H-\tfrac{1}{2}}\sqrt{E[v_s]}\mathrm{d}s\, E[v_t]\mathrm{d}t$ and $J(\theta) = \tfrac{\left(1+2 \rho^2\right) \eta^2 H + 4\rho^2 \eta^2 H(H+1) B(H+3 / 2, H+3 / 2)}{8(H+\tfrac{1}{2})^2( H+1)}$ with beta function $B(\cdot).$}
    \label{tab:models}
  \end{minipage}%
}
\end{center}

\section{Distribution-based Calibration}
\label{calibration}
In this section, we propose a model-free calibration methodology based on our previous results. Section~\ref{cal:experiment} carries out the calibration on real S\&P~500 data, and Section~\ref{cal:mot} discusses how the resulting marginal family feeds into path-dependent applications.

\subsection{Calibration Experiment}
\label{cal:experiment}
In the following, we approximate the bounded functions $\kappa_2(\theta)$, $\kappa_3(\theta)$, and $\kappa_4(\theta)$ as constants for tractability. We then regress the parameters $\{\kappa_2, \kappa_3, \kappa_4, \beta_1, \beta_2, m, n\}$ on market-implied volatility points based on the implied volatility approximation~\eqref{imp vol} in Theorem~\ref{Theorem 2}. The calibrated parameters can be interpreted as follows: $\kappa_2$, $\kappa_3$, $\kappa_4$ are multipliers of the cumulants of the standardized log return, $\beta_1$, $\beta_2$ are the decay rates of the high-order cumulants, and $m$, $n$ are Edgeworth-adjustment parameters that classify the marginal distribution type of the log return (see Section~\ref{sec:main-framework} and Table~\ref{tab:models}). In addition, we can recover the log return density from \textnormal{Eq.}~\eqref{q}, which is analytical and enables explicit computation under derivatives and integration.

The calibration design above is characterized by two features that distinguish it from established calibration practices. First, compared to model-based calibrations such as Heston~(\cite{heston1993closed}) and SABR~(\cite{hagan2002managing}), which commit to specific local- or stochastic-volatility dynamics whose parameters absorb the misspecification risk of the assumed SDE, our calibration is model-free and produces time-independent parameters: a single set $\{\kappa_2,\kappa_3,\kappa_4,\beta_1,\beta_2,m,n\}$ is fit jointly across the entire short-maturity surface from the asymptotic expansion~\eqref{imp vol} under condition~\eqref{density convergence}, with no assumption on the dynamics of the underlying price or its volatility. The only modeling assumption is the cumulant-level condition~\eqref{density convergence} on the log-return marginals, rather than a process-level one.

Second, our method calibrates directly using the implied-volatility surface. Compared with SVI~(\cite{gatheral2004parsimonious,gatheral2014arbitrage}), our framework calibrates through an asymptotically derived expansion under condition~\eqref{density convergence}. SVI is a purely geometric per-slice shape fit with no distributional meaning and no recoverable density, whereas in our framework the parameters carry direct distributional meaning and an explicit log-return density follows from~\eqref{q}, ensuring asymptotic arbitrage-free consistency while committing only to distributional conditions on the marginals.

\begin{figure}[!htbp]
    \centering
 \begin{subfigure}[b]{0.85\textwidth}
        \centering
        \includegraphics[width=\textwidth]{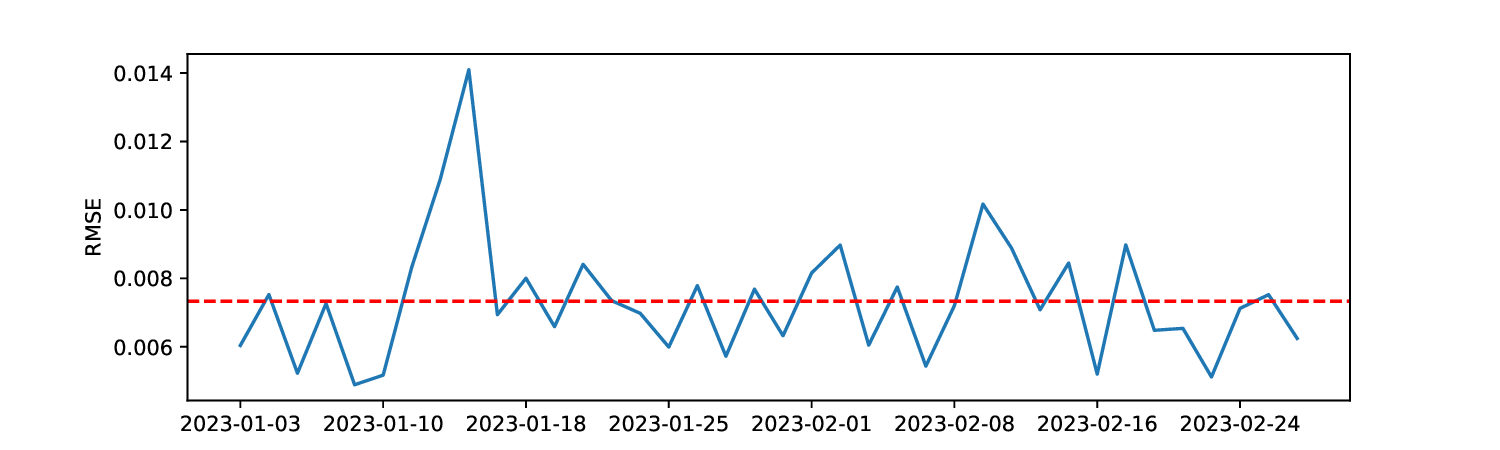}
    \end{subfigure}
    \hfill
    \begin{subfigure}[b]{0.85\textwidth}
        \centering
        \includegraphics[width=\textwidth]{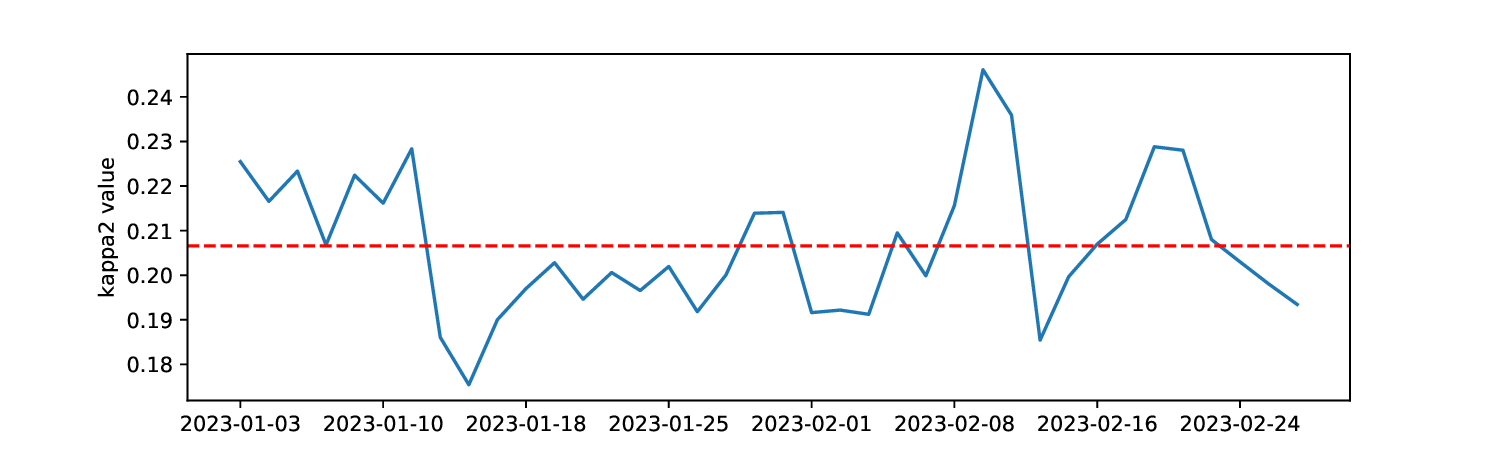}
    \end{subfigure}
\caption{The daily RMSE (top) and $\kappa_2$ values (down) of the distribution-based calibrations to SPX options from 2023-01-03 to 2023-02-28. Details of the data processing method follow Appendix \ref{data process}. The dashed red lines represent average level.}
\label{cal_rmse}
\end{figure}

Following this method, we perform daily distribution-based calibrations to S\&P 500 (SPX) index options data from 2023-01-03 to 2023-02-28. We use the data-filtering method in Appendix \ref{data process}. Importantly, we calibrate market-implied volatility points $\sigma(\sqrt{\theta}z, \theta)$ within the range $z \in [-0.75, 0.75]$ and $\theta \in [3/365, 90/365].$ Finally, we obtain a daily average of 540 option quotes.
\begin{figure}[!htbp]
    \centering
 \begin{subfigure}[b]{0.85\textwidth}
        \centering
        \includegraphics[width=\textwidth]{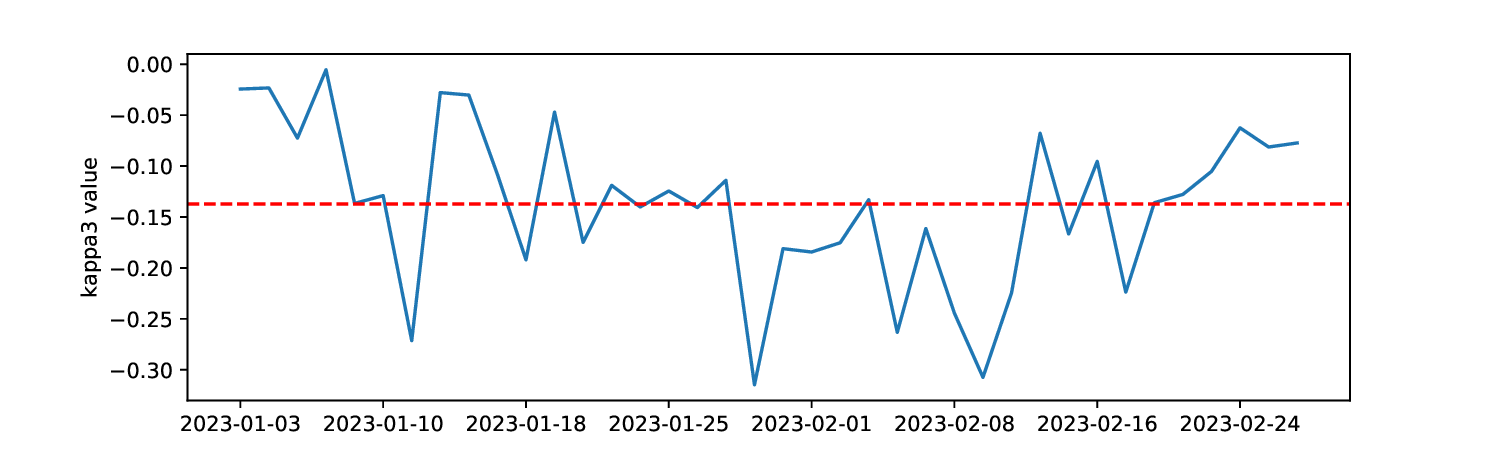}
    \end{subfigure}
    \hfill
    \begin{subfigure}[b]{0.85\textwidth}
        \centering
        \includegraphics[width=\textwidth]{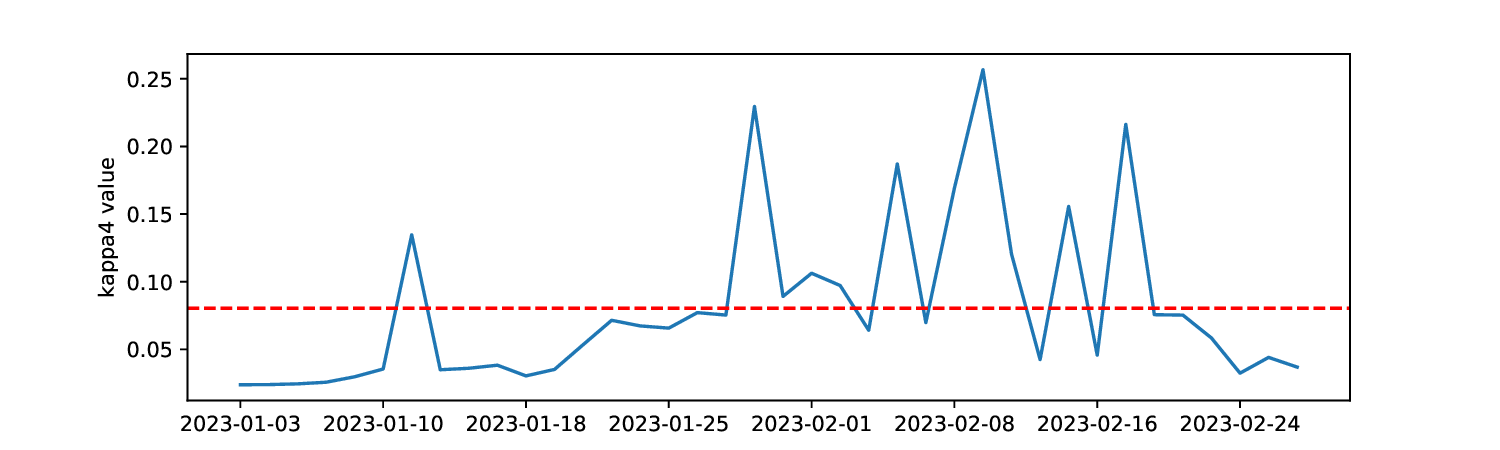}
    \end{subfigure}
\caption{The daily $\kappa_3$ (top) and $\kappa_4$ values (down) of the distribution-based calibrations to SPX options from 2023-01-03 to 2023-02-28. Details of the data processing method follow Appendix \ref{data process}. The dashed red lines represent the average level.}
\label{cal_kappa3}
\end{figure}
We perform an optimization over $\Theta = \{\kappa_2, \kappa_3, \kappa_4, \beta_1, \beta_2, m, n\}$ by minimizing the root-mean-square error (RMSE) between approximated and market-implied volatility points. We set the initial guess of the parameters as $\kappa_2 = 0.3$, $\kappa_3 = 0$, $\kappa_4 = 1$, $\beta_1 = 0.5$, $\beta_2 = 1$, $m = 1$, $n = 0$, where $\beta_1$, $\beta_2$, $m$ and $n$ are set based on a regular SVM guess. The purpose here is to make the calibration results more convincing if we observe a consistent deviation of the calibrated parameters from the initial guess.

Figure \ref{cal_rmse} plots the daily RMSE and $\kappa_2$ (representing the volatility level) values of the calibration results. We obtain an average of $0.0073$ RMSE value, which is relatively small (3.5\%) compared to the average volatility level of $\bar \kappa_2 = 0.206$. The values of RMSE also exhibit a robust pattern around the average level, except for a peak value of $0.014$. Overall, such good and consistent performance under a limited number of parameters demonstrates the effectiveness of our proposed distribution-based calibration method.

Figure \ref{cal_kappa3} shows the daily calibrated values of $\kappa_3$ (representing the skewness multiplier) and $\kappa_4$ (representing the kurtosis multiplier). The log return distributions all exhibit negative skewness with an average of $-0.137$, which is consistent with the market's typical behavior. And the kurtosis values are mostly positive and have an average of $0.080$. In comparison, a Black-Scholes model has $\kappa_4 = 0$. Overall, the calibrated log return distributions are negatively skewed and mostly heavy-tailed.

\begin{figure}[!htbp]
    \centering
 \begin{subfigure}[b]{0.85\textwidth}
        \centering
        \includegraphics[width=\textwidth]{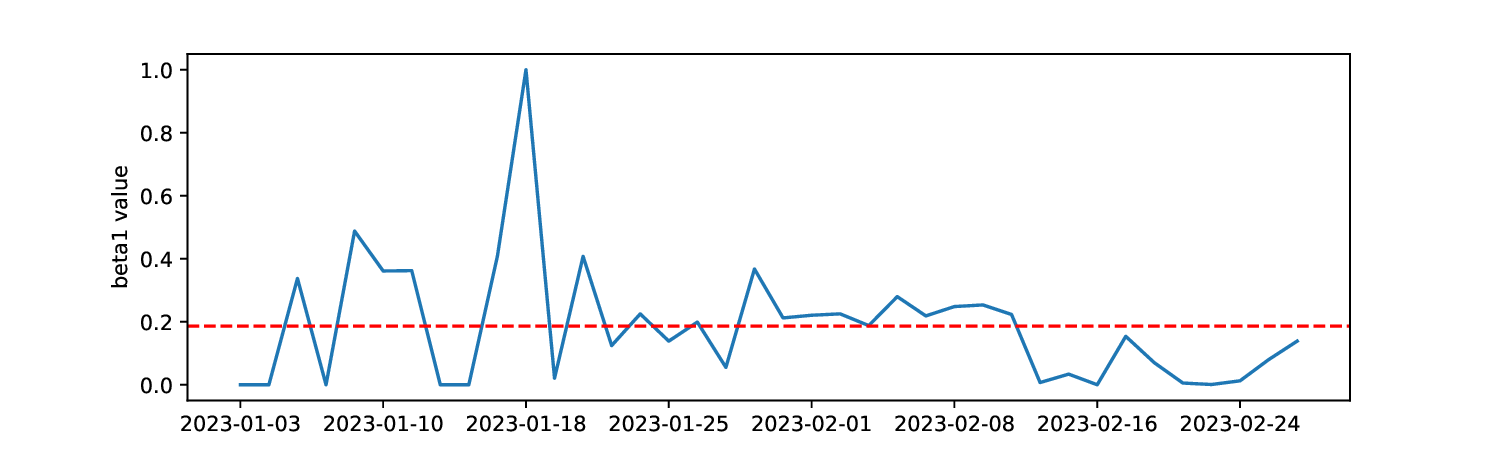}
    \end{subfigure}
    \hfill
    \begin{subfigure}[b]{0.85\textwidth}
        \centering
        \includegraphics[width=\textwidth]{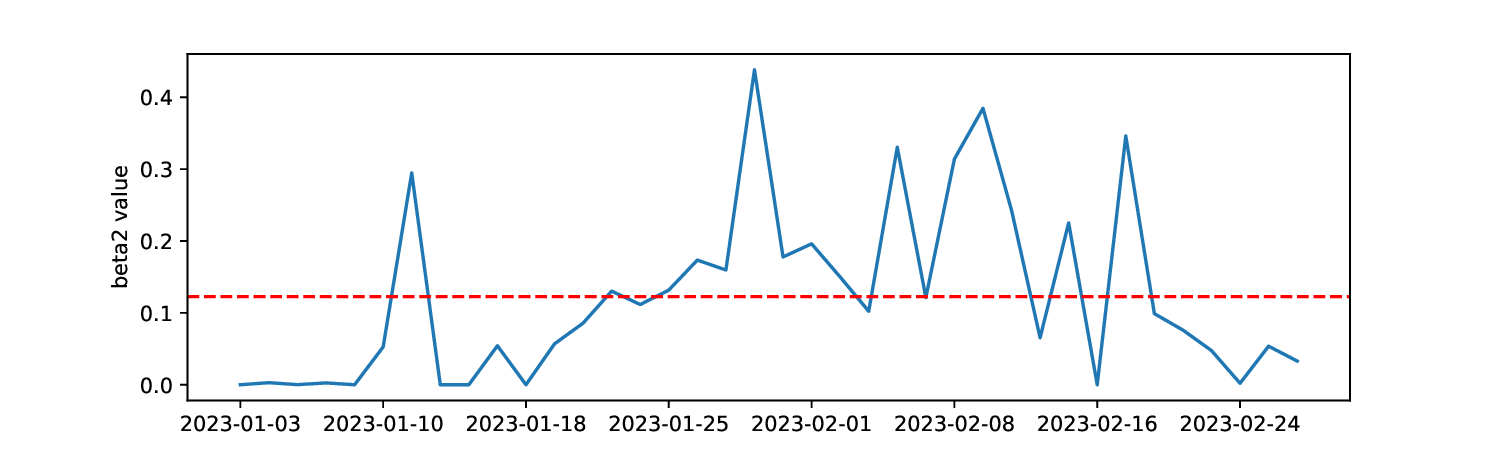}
    \end{subfigure}
\caption{The daily $\beta_1$ (top) and $\beta_2$ (bottom) values of the distribution-based calibrations to SPX options from 2023-01-03 to 2023-02-28. Details of the data processing method follow Appendix \ref{data process}. The dashed red lines represent the average level.}
\label{cal_beta1}
\end{figure}

Next, Figure \ref{cal_beta1} shows the daily calibrated values of $\beta_1$ (representing the decay rate of skewness) and $\beta_2$ (representing the decay rate of kurtosis). We see from the figure that $\beta_1$ values mostly range from $0$ to $0.5$ with an average of $0.186$, which coincides with the well-known empirical observation of a rough volatility. Moreover, decay rate $\beta_1$ is stable except for a peak value of $1$. Likewise, decay rate $\beta_2$ is also stable and ranges from $0$ to $0.4$. The values of $\beta_1$ and $\beta_2$ can occasionally drop to zero (we set a lower bound of zero for the decay rates in the optimization because it is required in our assumption). Such zero values are consistent with the empirical observations in Figure \ref{empirical}, as some $\beta_1$ and $\beta_2$ estimates would otherwise be negative, which would violate our regularity assumptions. 

\begin{figure}[!htbp]
    \centering
 \begin{subfigure}[b]{0.85\textwidth}
        \centering
        \includegraphics[width=\textwidth]{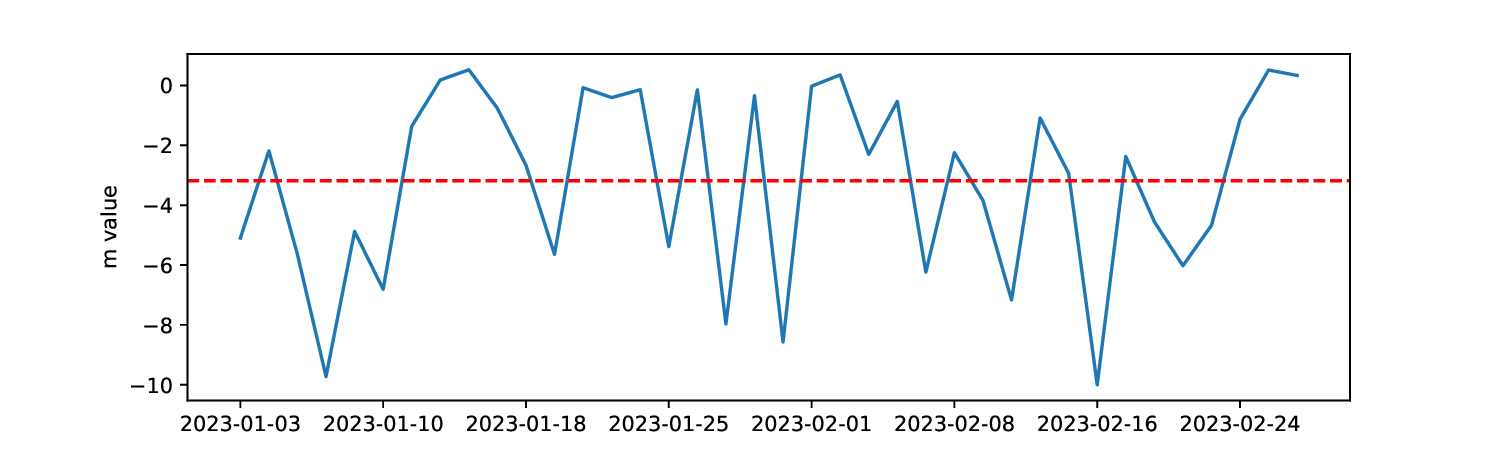}
    \end{subfigure}
    \hfill
    \begin{subfigure}[b]{0.85\textwidth}
        \centering
        \includegraphics[width=\textwidth]{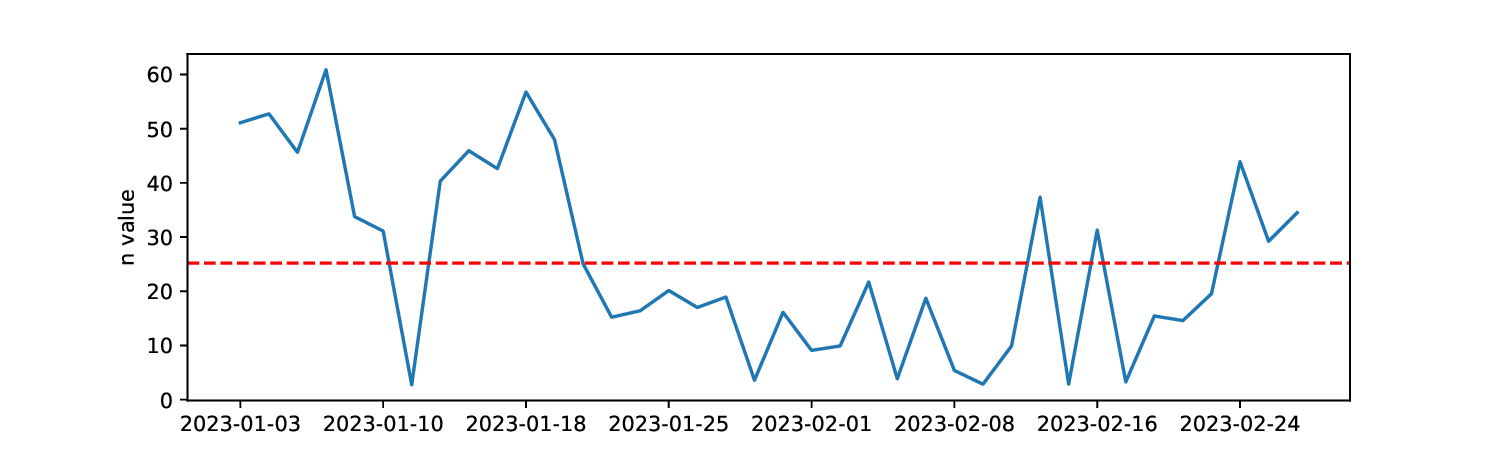}
    \end{subfigure}
\caption{The daily distribution type parameters $m$ (top) and $n$ values (down) of the distribution-based calibrations to SPX options from 2023-01-03 to 2023-02-28. Details of the data processing method follow Appendix \ref{data process}. The dashed red lines represent the average level.}
\label{cal_m}
\end{figure}

We next analyze the distribution-type parameters $m$ (characterizing the type of skewness) and $n$ (characterizing the type of kurtosis) via Figure~\ref{cal_m}. Calibrated values of $m$ are negative and stable, ranging mostly from $-10$ to $0$ with an average of $-3.18$, distinct from both the SVM class ($m=1$) and the pure distribution-based class ($m=0$); this locates the SPX log-return distribution in a class with stronger Edgeworth adjustment than either standard family, and indicates that fitting the market with a model anchored to $m=1$ would systematically misprice the contribution of skewness to the ATM skew. Values of $n$ are positive but considerably more volatile, consistent with the sub-leading role of $n$ in the expansion~\eqref{imp vol}: the $n$-bearing term decays the fastest, making $n$ the least identifiable component of the calibration. This low identifiability suggests that in practice one may fix $n$ at a representative value (e.g., $n=0$ or $n=2$ as practical choices) and calibrate the remaining six parameters.

\subsection{Toward Path-dependent Applications via Martingale Optimal Transport}
\label{cal:mot}
The calibration carried out in Section~\ref{cal:experiment} fits the IV surface but stops at the marginal level; this subsection sketches how the calibrated marginals can be used for path-dependent applications.

Beyond the validation role played above, the calibrated marginal family $\{q_\theta^{(m,n)}\}_{\theta\in(0,1]}$ is a self-contained input for path-dependent financial tasks. Two such tasks deserve emphasis: robust pricing and robust hedging of exotic options, both of which can be approached through the theory of optimal embedding or, more generally, martingale optimal transport (MOT); see, e.g., \cite{beiglbock2013model,dolinsky2014martingale}.

Specifically, once the marginal family is fixed, MOT computes the supremum and infimum of the expected payoff of a path-dependent option over all martingale couplings consistent with these marginals, yielding sharp model-independent bounds on exotic prices; explicit constructions of consistent martingales are available~(\cite{madan2002making}). By contrast, process-based calibrations (local volatility, SVMs, jump-diffusion) follow a different route: once an SDE is assumed and calibrated, hedging ratios and exotic prices are obtained as precise point values from the dynamics, but these values rely entirely on the assumed SDE being correct, which is an assumption that European option prices alone do not pin down. Our calibration paired with MOT thus offers an alternative: a range of values derived from the calibrated marginals, with no commitment to a particular SDE.

The robust hedging problem can be cast in optimal-control form as follows. Let $G(S_{\cdot})$ be the payoff of a path-dependent option depending on the price path $(S_t)_{t\in[0,1]}$ on a finite set of monitoring dates $0=t_0<t_1<\cdots<t_N=1$, with $S_{t_i} = S_0\,e^{Z_{t_i}}$. Let $\mathcal{M}(\{q_\theta^{(m,n)}\})$ denote the set of martingale measures $Q$ under which $(S_t)$ is a martingale and the marginal law of $S_{t_i}$ agrees with the calibrated marginal at maturity $\theta = t_i$ for each $i$. The super-replication cost is
\begin{equation}
\label{robust hedging}
\bar{V}_0 \;=\; \inf_{(H,c)\in\mathcal{A}}\;\Bigl\{\,c \;:\; c + \sum_{i=0}^{N-1} H_{t_i}\bigl(S_{t_{i+1}} - S_{t_i}\bigr) \;\ge\; G(S_{\cdot})\;\;Q\text{-a.s. for every } Q\in\mathcal{M}(\{q_\theta^{(m,n)}\})\,\Bigr\},
\end{equation}
where $\mathcal{A}$ is the set of admissible pairs of an initial capital $c\in\mathbb{R}$ and a predictable hedging strategy $(H_{t_i})_{i=0}^{N-1}$. The control variable here is the dynamic position $H$ in the underlying. The sub-replication cost $\underline{V}_0$ is defined analogously with $\inf$ replaced by $\sup$ and the inequality reversed. By the MOT duality, the value~\eqref{robust hedging} equals $\sup_{Q\in\mathcal{M}(\{q_\theta^{(m,n)}\})}E^{Q}[G(S_{\cdot})]$, so the same calibrated marginals deliver both the optimal super-replication cost and the corresponding price bound.

In summary, the distribution-based calibration developed in Section~\ref{calibration} effectively captures the market's features (the log return's skewness, kurtosis, decay patterns, and distribution type) with a parsimonious setting and parameters of direct distributional meaning. The calibrated marginal family $\{q_\theta^{(m,n)}\}$ is the natural primitive for the path-dependent tasks formalized above: robust hedging via~\eqref{robust hedging}, robust pricing of exotic options through the associated MOT bounds, and risk evaluations that respect the marginal constraints without committing to any specific SDE. The calibration is therefore a self-contained first step toward these applications, the systematic numerical development of which we leave to future work.

\section{Conclusion}
\label{conclusion}
This study introduces a unified, distribution-based framework for analyzing short-term implied volatility asymptotics. The central claim is that short-time ATM asymptotics are governed by the scaling rates of low-order cumulants of the marginal log-return distribution, specifically the pair $(\beta_1, \beta_2)$ controlling how fast skewness and kurtosis decay to zero, together with a small set of classification parameters $(m,n)$ that capture the type of the marginal density. From this viewpoint, roughness ($H < 1/2$), jumps, and decaying leverage are different mechanisms that realize particular values of $(\beta_1, \beta_2; m, n)$.

The framework accommodates arbitrary decay rates for high-order cumulants, enabling a far broader range of asymptotic regimes than conventional models. For instance, it identifies novel regimes where kurtosis dominates skewness in shaping the ATM skew, a consequence of the free $(\beta_1, \beta_2)$ pair, and where the curvature-skew explosion ratio exceeds $2{:}1$, as observed empirically in S\&P~500 data. In our framework, the celebrated short-time power law $\partial_k\sigma_{\mathrm{BS}}(0,\theta) \sim \theta^{H-1/2}$ corresponds to the specific class $(m,n)=(1,0)$ of a more general cumulant-scaling phenomenon; traditional rough and regular volatility models (e.g., rough Bergomi) fail to replicate the empirical explosion ratio due to their rigid $\beta_2 = 2\beta_1$ constraint. This perspective does not diminish the empirical importance of the rough-volatility discovery; rather, it reframes it. The rough-volatility literature identified a specific process mechanism that generates the observed power law, and our framework shows that this power law is a characteristic outcome of a more general cumulant-scaling regime which multiple process mechanisms can realize.

The framework's model-independence enables the unification of diverse volatility modeling paradigms under a single theoretical structure. By requiring only cumulant decay rates as inputs, it integrates SVMs (e.g., Heston model, 3/2 volatility model), rough volatility models (e.g., rough Bergomi model), and distribution-based models into a coherent analytical system via the $(m,n)$ classification. The classification itself carries meaning: the calibrated $(m,n)$ values from SPX data (Section~\ref{calibration}) suggest that the S\&P~500 market belongs to a class distinct from both the SVM family ($m=1$) and the distribution-based family ($m=0$), offering a quantitative diagnostic for model selection.

The distribution-based calibration developed in Section~\ref{calibration} achieves two goals. First, it validates the IV expansion on real S\&P~500 data and shows that the empirically calibrated $(\beta_1,\beta_2)$ and $(m,n)$ behave consistently across the panel, confirming the theoretical regime predictions. Second, the calibrated marginal family $\{q_\theta^{(m,n)}\}$ is itself the natural primitive for path-dependent financial applications, as developed in Section~\ref{cal:mot}: robust hedging, formalized as the optimal-control problem~\eqref{robust hedging}, and the associated MOT bounds on exotic prices and super-/sub-replication costs, all without committing to any specific SDE.

Our framework therefore does not compete with the rough-volatility, jump, or decaying-leverage models in the process-based literature: those models specify particular mechanisms realizing a given cumulant-scaling regime, whereas ours abstracts from these process-level specifications and provides a common diagnostic and calibration interface for them at the level of the marginal distribution. The numerical implementation of the robust-hedging problem~\eqref{robust hedging} formulated in Section~\ref{cal:mot}, and more broadly the development of MOT-based path-dependent valuation on top of the calibrated marginals, is a natural direction for future work.

\appendix
\section{Proofs in Section \ref{sec 2}}
\subsection{Proof of Theorem \ref{Th1}}
\label{proof Th1}
\begin{proof}
Let $\alpha > 1$ be the constant that ensures Condition \eqref{density convergence}. By the Cauchy-Schwarz inequality, we have, for $\epsilon \in (0, 2\alpha - 2),$
\begin{align*}
&  \int_{-\infty}^z \int_{-\infty}^\zeta |p_\theta(x) - q_\theta^{(m,n)}(x)| \, dx \, e^{\sigma_0 \zeta} \mathrm{d}\zeta \\
\leq &  \int_{-\infty}^z \sqrt{\int_{-\infty}^\zeta \tfrac{1}{(1 + x^2)^{2\alpha - 1/2 - \epsilon}} \mathrm{d}x}  \sqrt{\int_{-\infty}^\zeta (1 +x^2)^{2\alpha - 1/2 -\epsilon}|p_\theta(x) - q_\theta^{(m,n)}(x)|^2 \mathrm{d}x} \, e^{\sigma_0\zeta} \mathrm{d}\zeta \\
\leq & e^{\sigma_0 z} \sup_{x\leq z_0} (1 + x^2)^\alpha |p_\theta(x) - q_\theta^{(m,n)}(x)|\\
& \quad \quad \quad \cdot \int_{-\infty}^\infty (1+x^2)^{-1/2-\epsilon} \mathrm{d}x \int_{-\infty}^z \sqrt{\int_{-\infty}^\zeta \tfrac{1}{(1 + x^2)^{2\alpha - 1/2 - \epsilon}} \mathrm{d}x} \, \mathrm{d}\zeta\\
= & e^{\sigma_0 z} \sqrt{\tfrac{\sqrt{\pi}\Gamma(\epsilon)}{\Gamma(\epsilon + 1/2)}}  \sup_{x\leq z_0} (1 + x^2)^\alpha |p_\theta(x) - q_\theta^{(m,n)}(x)|\int_{-\infty}^z \sqrt{\int_{-\infty}^\zeta \tfrac{1}{(1 + x^2)^{2\alpha - 1/2 - \epsilon}} \mathrm{d}x} \, \mathrm{d}\zeta \\
=& o(\theta^{\bar{\beta}_\epsilon})
\end{align*}
uniformly for $z \leq z_0$, where $2\alpha - 1/2 - \epsilon > 3/2$ ensures integrability. Combining with Eq. \eqref{put price}, the put option price admits the representation:
\begin{equation*}
\tfrac{P(F e^{\sigma_0 z}, \theta)}{F  e^{-r\theta} \sigma_0} = \int_{-\infty}^z \int_{-\infty}^{\zeta} q_\theta^{(m,n)}(x) \, \mathrm{d}x \, e^{\sigma_0 \zeta} \mathrm{d}\zeta + o(\theta^{\bar{\beta}_\epsilon})
\end{equation*}
uniformly for all $z \leq z_0$. 

Recall the integral operator $\mathrm{I}^2$ defined in \textnormal{Eq.}~\eqref{pricing op} of Section~\ref{sec:main-framework}.
Integration by parts yields 
$$\mathrm{I}^2 \phi(z + \tfrac{\sigma_0}{2}) = \tfrac{1}{\sigma_0}\left(\Phi(z+ \tfrac{\sigma_0}{2})e^{\sigma_0 z} - \Phi(z- \tfrac{\sigma_0}{2})\right),$$
\begin{align*}
& \mathrm{I}^2 \left(\phi\left(z+ \tfrac{\sigma_0}{2}\right)\kappa_3 \left(H_3(x+\tfrac{\sigma_0}{2}) - m\sigma_0 H_2\left(x+ \tfrac{\sigma_0}{2}\right)\right)\theta^{\beta_1}\right) \\
=& \phi\left(z + \tfrac{\sigma_0}{2}\right)\kappa_3(H_1\left(x+ \tfrac{\sigma_0}{2}) + (1-m)\sigma_0\right)e^{\sigma_0 z}\theta^{\beta_1}+ (m-1)\sigma_0^2 \Phi\left(z- \tfrac{\sigma_0}{2}\right)\theta^{\beta_1},
\end{align*}
$$
\mathrm{I}^2 \left(\tfrac{\kappa_3^2}{2}\phi(z) H_6(z)\theta^{2\beta_1}\right) = \tfrac{\kappa_3^2}{2} \phi(z) H_4(z)\theta^{2\beta_1} + O(\theta^{2\beta_1 +\tfrac{1}{2}}),
$$
and 
\begin{align*}
& \mathrm{I}^2\left(\phi\left(z +\tfrac{\sigma_0}{2}\right)\kappa_4 \left(H_4\left(z + \tfrac{\sigma_0}{2}\right) - n\sigma_0 H_3\left(z+\tfrac{\sigma_0}{2}\right)\right)e^{\sigma_0z}\theta^{\beta_2}\right) \\
=& \phi\left(z + \tfrac{\sigma_0}{2}\right)\kappa_4\left(H_2\left(z+ \tfrac{\sigma_0}{2}\right) + (1-n) \sigma_0 H_3\left(z+ \tfrac{\sigma_0}{2}\right)\right)\theta^{\beta_2} + O(\theta^{\beta_2 + 1}).
\end{align*}
We recall the definition of $P^{(m,n)}(z, \theta)$. Since $\bar\beta_{\epsilon} <\min\{2\beta_1 + \tfrac{1}{2}, \beta_2+ 1\}$, incorporating the higher-order terms $O(\theta^{2\beta_1 + \tfrac{1}{2}})$ and $O(\theta^{\beta_2 + 1})$ into the error term $o(\theta^{\bar{\beta}_\epsilon})$ yields the desired result.
\end{proof}

\subsection{Proof of Theorem \ref{Theorem 2}}
\label{proof Th2}
\begin{proof}
    Fix $z \in \mathbb{R}$. Define the normalized put price function:
\[
P_\theta(\sigma) := \tfrac{P_\text{BS
}(F e^{\sqrt{\theta} z}, \theta, \sigma)}{F e^{-r\theta} \sqrt{\theta}} := \tfrac{1}{\sqrt{\theta}} \left( \Phi\left( \tfrac{z}{\sigma} + \tfrac{\sigma \sqrt{\theta}}{2} \right) e^{\sqrt{\theta} z} - \Phi\left( \tfrac{z}{\sigma} - \tfrac{\sigma \sqrt{\theta}}{2} \right) \right).
\]
We see that
\[
P_\theta \colon [0, \infty] \to \left[ \tfrac{(e^{\sqrt{\theta} z} - 1)_+}{\sqrt{\theta}}, \tfrac{e^{\sqrt{\theta} z}}{\sqrt{\theta}} \right]
\]
is a strictly increasing function. We have from Theorem \ref{Th1} that \[
\begin{aligned}
\tfrac{P(F e^{\sqrt{\theta} z}, \theta)}{F e^{-r\theta} \sqrt{\theta}}& = P_{\theta}(\kappa_2) + \kappa_2 \kappa_3 \phi\left(\tfrac{z}{\kappa_2} + \tfrac{\kappa_2 \sqrt{\theta}}{2}\right) H_1\left(\tfrac{z}{\kappa_2} + \tfrac{\kappa_2 \sqrt{\theta}}{2}\right) e^{\sqrt{\theta} z} \theta^{\beta_1}\\
&\quad + \kappa_2\kappa_4\phi(\tfrac{z}{\kappa_2})H_2(\tfrac{z}{\kappa_2})\theta^{\beta_2} + o(\theta ^{\beta_1 \wedge \beta_2})
\\
&= P_{\theta}(\kappa_2) + O(\theta^{\beta_1\wedge \beta_2}).
\end{aligned}
\]

Therefore, the implied volatility \(\sigma_{\text{BS}}\) satisfies:
\[
\sigma_{\text{BS}}(\sqrt{\theta} z, \theta) = P_{\theta}^{-1}\left(P_{\theta}(\kappa_2) + O(\theta^{\beta_1 \wedge \beta_2})\right).
\]
By assumption, $\kappa_2$, as a function of $\theta$, is bounded by some $L > 0$. The function \(P_\theta\) converges as \(\theta \to 0\) to
\[P_0(\sigma) := z\Phi\left(\tfrac{z}{\sigma}\right) + \sigma\phi\left(\tfrac{z}{\sigma}\right)\]
pointwise, and by Dini's theorem, this convergence is uniform on \([0, L]\). Since the limit function \(P_0\) is strictly increasing, the inverse functions \(P_\theta^{-1}\) converges to \(P_0^{-1}\). Again by Dini's theorem, this convergence is uniform and in particular, $\{P_\theta^{-1}\}_{\theta \in (0,1)}$ are equicontinuous. Thus, we conclude that \(\sigma_{\text{BS}}(\sqrt{\theta}z, \theta) - \kappa_2 \to 0\) as \(\theta \to 0\).
Furthermore, since $P_{\theta}(\sigma_{\text{BS}}) = P_\theta(\kappa_2) + O(\beta^{\beta_1 \wedge \beta_2})$, the mean-value theorem yields $\sigma_{\text{BS}}(\sqrt{\theta}z, \theta) - \kappa_2 = O(\beta^{\beta_1 \wedge \beta_2}).$

Next, we note that $$\tfrac{\mathrm{d}}{\mathrm{d}\sigma}P_\theta(\sigma) = \phi(\tfrac{z}{\sigma} - \tfrac{\sigma\sqrt{\theta}}{2}),\quad \tfrac{\mathrm{d}^2}{\mathrm{d}\sigma^2} P_\theta(\sigma) = \left(\tfrac{z^2}{\kappa_2^3} - \tfrac{\kappa_2\theta}{4}\right)\phi(\tfrac{z}{\kappa_2} - \tfrac{\kappa_2\sqrt{\theta}}{2}).$$
By the result of Theorem \ref{Th1},
$$\begin{aligned}
A_0 & := \kappa_2\phi(\tfrac{z}{\kappa_2} - \tfrac{\kappa_2\sqrt{\theta}}{2}) \left(\kappa_3  \left(H_1(\tfrac{z}{\kappa_2}+\tfrac{\kappa_2\sqrt{\theta}}{2})+ (1-m)\sigma_0 \right) \theta^{\beta_1} \right.\\
& \quad \quad \quad \left.+ \kappa_2\kappa_4 \left(H_2(\tfrac{z}{\kappa_2}+ \tfrac{\kappa_2\sqrt{\theta}}{2}) + (1-n) \sigma_0 H_1(\tfrac{z}{\kappa_2}+ \tfrac{\kappa_2\sqrt{\theta}}{2})\right)\theta^{\beta_2}\right)\\
&= P_\theta(\sigma_{\text{BS}}) - P_\theta(\kappa_2) + o(\theta^{\beta_1\wedge \beta_2})\\
&= \phi(\tfrac{z}{\kappa_2} - \tfrac{\kappa_2\sqrt{\theta}}{2})(\sigma_{\text{BS}} - \kappa_2) + o(\theta^{\beta_1 \wedge\beta_2}),
\end{aligned}$$
where the last equality follows from the first-order Taylor expansion and the previous result that $\sigma_{\text{BS}} - \kappa_2 = O(\theta^{\beta_1 \wedge \beta_2})$. This result then leads to the approximation 
\begin{equation}
\label{1-order}\sigma_{\text{BS}} = \kappa_2\left(1 + \kappa_3\left(\tfrac{z}{\kappa_2} + (\tfrac{3}{2}-m)\kappa_2\sqrt{\theta}\right)\theta^{\beta_1} + \kappa_4\left(\tfrac{z^2}{\kappa_2^2} - 1\right) \theta^{\beta_2}\right) + o(\theta^{\beta_1 \wedge\beta_2}).
\end{equation}

Finally, we increase the asymptotic accuracy to $O(\theta^{\bar{\beta}_\epsilon})$. Let 
\begin{equation}
\label{sigma_BS}\sigma_{\text{BS}}(\sqrt{\theta}z, \theta)\equiv \kappa_2 + A_1 \theta^{\beta_1} + A_2 \theta^{\beta_2} + A_3,\end{equation} where $A_1$ and $A_2$ are given from \textnormal{Eq.} \eqref{1-order} as
$$A_1 = \kappa_2 \kappa_3 \left(\tfrac{z}{\kappa_2} + (\tfrac{3}{2}-m)\kappa_2\sqrt{\theta}\right), \quad A_2= \kappa_2\kappa_4\left(\tfrac{z^2}{\kappa_2^2} - 1\right).$$ 
From Theorem \ref{Th1} and the second-order Taylor expansion,
$$\begin{aligned}&A_0 + \tfrac{1}{2}\kappa_2\kappa_3\phi(\tfrac{z}{\kappa_2})H_4(\tfrac{z}{\kappa_2})\theta^{2\beta_1} + (m-1) \kappa_2\kappa_3 \sigma_0^2\Phi(z - \tfrac{\sigma_0}{2})\theta^{\beta_1}\\
            = & P_{\theta}(\sigma_{\text{BS}}) - P_{\theta}(\kappa_2) + o(\theta^{\bar{\beta}_\epsilon})\\
            = & \phi(\tfrac{z}{\kappa_2}-\tfrac{\kappa_2\sqrt{\theta}}{2})(\sigma_{\text{BS}}-\kappa_2) + \tfrac{1}{2}\left(\tfrac{z^2}{\kappa_2^3} - \tfrac{\kappa_2\theta}{4}\right)\phi(\tfrac{z}{\kappa_2} - \tfrac{\kappa_2\sqrt{\theta}}{2})(\sigma_{\text{BS}}- \kappa_2)^2 + o(\theta^{\bar{\beta}_\epsilon})\\
            =& \phi(\tfrac{z}{\kappa_2}-\tfrac{\kappa_2\sqrt{\theta}}{2})\left((A_1 \theta^{\beta_1} + A_2 \theta^{\beta_2} + A_3) + \tfrac{z^2}{2\kappa_2^3}A_1^2\theta^{2\beta_1}\right) + o(\theta^{\bar{\beta}_\epsilon}),
            \end{aligned}$$
from which
        $$A_3 = \kappa_3^2\left(-\tfrac{3}{\kappa_2}z^2 + \tfrac{3\kappa_2}{2}\right)\theta^{2\beta_1}  + (2- n)\kappa_2\kappa_4 z\theta^{\beta_2 + 1/2} +(m-1)\kappa_2^3\kappa_3\tfrac{\Phi(\tfrac{z}{\kappa_2} - \tfrac{\kappa_2}{2})}{\phi(\tfrac{z}{\kappa_2} - \tfrac{\kappa_2}{2})} \theta ^{\beta_1 + 1}+ o(\theta^{\bar{\beta}_\epsilon}).$$
And the result follows from substituting $A_1$, $A_2$, and $A_3$ into \textnormal{Eq.} \eqref{sigma_BS}.
\end{proof}

\subsection{Proof of Theorem \ref{ATM skew}}
\label{proof Th3}
\begin{proof}
    From known results (cf. \cite{fukasawa2012normalizing}), the derivatives of implied volatility satisfy:
    \begin{equation}
    \label{formula skew}
\begin{aligned}
\partial_k \sigma_{\text{BS}}(k, \theta) &= \tfrac{Q(k \geq \sigma_0X_\theta) - \Phi(f_2(k, \theta))}{\sqrt{\theta} \phi(f_2(k, \theta))},\\
\partial_k^2 \sigma_{\text{BS}}(k, \theta) &= \tfrac{p_\theta(\tfrac{k}{\sigma_0})}{\sigma_0 \sqrt{\theta} \phi(f_2(k, \theta))} - \sigma_{\text{BS}}(k, \theta) \partial_k f_1(k, \theta) \partial_k f_2(k, \theta),
\end{aligned}
\end{equation}
where
\[
\begin{aligned}
f_1(k, \theta) &= \tfrac{k}{\sqrt{\theta} \sigma_{\text{BS}}(k, \theta)} - \tfrac{\sqrt{\theta} \sigma_{\text{BS}}(k, \theta)}{2}, \\
f_2(k, \theta) &= \tfrac{k}{\sqrt{\theta} \sigma_{\text{BS}}(k, \theta)} + \tfrac{\sqrt{\theta} \sigma_{\text{BS}}(k, \theta)}{2}.
\end{aligned}
\]

\noindent\textbf{Step 1: Computation of ATM skew} 

At $k=0$, we observe from the definition of $q^{(m,n)}_\theta(x)$ and Condition \eqref{density convergence} that the distribution function:
\begin{equation*}
\begin{aligned}
Q(X_\theta \leq 0)& = \Phi\left(\tfrac{\sigma_0}{2}\right) + \kappa_3 \phi\left(\tfrac{\sigma_0}{2}\right)\left(1 - \tfrac{1}{4}\sigma_0^2 + \tfrac{m}{2}\sigma_0^2\right)\theta^{\beta_1} + \left(\tfrac{3}{2}-n\right)\sigma_0\kappa_4 \phi\left(\tfrac{\sigma_0}{2}\right)\theta^{\beta_2} + o(\theta^{\bar{\beta}_\epsilon})\\
& =  \Phi\left(\tfrac{\sigma_0}{2}\right) + \kappa_3 \phi\left(\tfrac{\sigma_0}{2}\right)\theta^{\beta_1} + \left(\tfrac{3}{2}-n\right)\sigma_0\kappa_4 \phi\left(\tfrac{\sigma_0}{2}\right)\theta^{\beta_2} + O(\theta^{\beta_1 + 1}) + o(\theta^{\bar{\beta}_\epsilon}).
\end{aligned}
\end{equation*}
And, from Theorem \ref{Theorem 2}, the expansion of implied volatility yields:
\begin{equation*}
f_2(0, \theta) =  \tfrac{\sqrt{\theta}\sigma_{\text{BS}}(0, \theta)}{2}= \tfrac{\kappa_2\sqrt{\theta}}{2}\left(1 - \kappa_4 \theta^{\beta_2} +\left(\tfrac{3}{2} -m\right) \kappa_2\kappa_3\theta^{\beta_1 +1/2}\right) + O(\theta^{\beta_1+3/2}) + o(\theta^{\bar{\beta}_\epsilon}).
\end{equation*}
This yields the Taylor expansion of the form
\begin{equation*}
\begin{aligned}
\Phi(f_2(0,\theta)) & = \Phi\left(\tfrac{\sigma_0}{2}\right) + \phi\left(\tfrac{\sigma_0}{2}\right)(f_2(0,\theta) - \tfrac{\sigma_0}{2}) + o(\theta^{\bar{\beta}_\epsilon})\\
&= \Phi\left(\tfrac{\sigma_0}{2}\right) - \phi\left(\tfrac{\sigma_0}{2}\right)\tfrac{\kappa_2\kappa_4}{2}\theta^{\beta_2+1/2} + O(\theta^{\beta_1+1}) + o(\theta^{\bar{\beta}_\epsilon}).
\end{aligned}.
\end{equation*}

Combining these results, we obtain, for any $\epsilon > 0$,
$$\theta^{-1/2} (Q(X_\theta \le 0) - \Phi(f_2(0,\theta))) = \phi(\tfrac{\sigma_0}{2})\left(\kappa_3 \theta^{\beta_1 - 1/2} + (2-n)\kappa_2\kappa_4 \theta^{\beta_2} \right)+ o(\theta^{\bar \beta^-_\epsilon}),$$
with $\bar \beta^-_\epsilon = \min\{\beta_1 + 1/2 -\epsilon, \bar{\beta}_\epsilon - 1/2\}$.
Finally, using 
\begin{equation}
    \label{phi diff}
    \begin{aligned}
    \phi(f_2(0,\theta)) &= \phi(\tfrac{\sigma_0}{2}) + \phi'(\tfrac{\sigma_0}{2})(f_2(0,\theta) - \tfrac{\sigma_0}{2})+ \theta o(f_2(0,\theta) - \tfrac{\sigma_0}{2})\\
    &=\phi(\tfrac{\sigma_0}{2}))(1 + O(\theta^{1 + \min\{\beta_1 + 1/2, \beta_2\}})),
    \end{aligned}
\end{equation}
we have from \textnormal{Eq.} \eqref{formula skew} that
\begin{equation*}
\partial_k \sigma_{\text{BS}}(0, \theta) = \kappa_3 \theta^{\beta_1 - 1/2} + (2-n)\kappa_2\kappa_4\theta^{\beta_2} + o(\theta^{\bar \beta^-_\epsilon}).
\end{equation*}

\noindent\textbf{Step 2: Computation of ATM curvature}

We analyze the two terms in the expression of ATM curvature (in \textnormal{Eq.} \eqref{formula skew}) separately:

\begin{itemize}
\item \textbf{Density term:} From \textnormal{Eq.} \eqref{q} and Condition \eqref{density convergence}, the density at zero has expansion:
$$\begin{aligned}p_\theta(0)&= \phi(\tfrac{\sigma_0}{2})\left(1 + (m-\tfrac{3}{2})\kappa_2 \kappa_3\theta^{\beta_1 + \tfrac{1}{2}}+\tfrac{1 -2m}{8}\kappa_2^3\kappa_3\theta^{\beta_1+ \tfrac{3}{2}} \right.\\
& \quad \quad \left.+ 3\kappa_4\theta^{\beta_2} - \tfrac{15}{2}\kappa_3^2\theta^{2\beta_1}\right) +o(\theta^{\bar{\beta}_\epsilon}),\end{aligned}$$
which, combined with \textnormal{Eq.} \eqref{phi diff}, further yields:
$$
\begin{aligned}\tfrac{p_\theta(0)}{\sigma_0 \sqrt{\theta} \phi(f_2(0, \theta))} &= \kappa_2^{-1}\theta^{-1}\left(1 + (m-\tfrac{3}{2})\kappa_2 \kappa_3\theta^{\beta_1 + \tfrac{1}{2}}\right.\\
& \quad +\left.\tfrac{1 -2m}{8}\kappa_2^3\kappa_3\theta^{\beta_1+ \tfrac{3}{2}} + 3\kappa_4\theta^{\beta_2} - \tfrac{15}{2}\kappa_3^2\theta^{2\beta_1}\right) + o(\theta^{\bar{\beta}_\epsilon -1}).
\end{aligned}$$

\item \textbf{Cross term:} $$\begin{aligned}
    \sigma_{\text{BS}}(0,\theta)\partial_k f_1(0, \theta) \partial_k f_2(0, \theta)
    = & \tfrac{1}{\theta \sigma_{\text{BS}}(0,\theta)} - \tfrac{\theta}{4} (\partial_k\sigma_{\text{BS}}(0,\theta))^2 \sigma_{\text{BS}}(0,\theta) \\
    = & \tfrac{1}{\theta \sigma_{\text{BS}}(0,\theta)} + O(\theta^{\bar{\beta}_\epsilon}),
\end{aligned}$$
where, according to Theorem \ref{Theorem 2},
$$\begin{aligned}
    \tfrac{1}{\theta \sigma_{\text{BS}}(0,\theta)}=& \kappa_2^{-1}\theta^{-1}\bigg(1 + (m-\tfrac{3}{2})\kappa_2\kappa_3\theta^{\beta_1 + \tfrac{1}{2}} - (m-1) \tfrac{\sqrt{2\pi}}{2} \kappa_2^2\kappa_3\theta^{\beta_1 + 1}\\
    & \quad \quad \quad - \tfrac{3\kappa_3^2}{2}\theta^{2\beta_1} + \kappa_4\theta^{\beta_2}\bigg) + o(\theta^{\bar{\beta}_\epsilon -1}).
\end{aligned}$$
\end{itemize}

Subtracting the two expansions above gives 
$$\begin{aligned}
    &\partial_k^2 \sigma_{\text{BS}}(0, \theta) \\=& \tfrac{p_\theta(0)}{\sigma_0 \sqrt{\theta} \phi(f_2(0, \theta))} - \sigma_{\text{BS}}(0, \theta) \partial_k f_1(0, \theta) \partial_k f_2(0, \theta)\\
    =& (\kappa_2 \theta)^{-1}\left(2\kappa_4\theta^{\beta_2}-6\kappa_3^2\theta^{2\beta_1} + \kappa_2^2\kappa_3((m-1)\tfrac{\sqrt{2\pi}}{2}\theta^{\beta_1+1} + \tfrac{1-2m}{8}\kappa_2\theta^{\beta_1 + \tfrac{3}{2}})\right)+ o(\theta^{\bar{\beta}_\epsilon -1})\\
    =& \tfrac{2\kappa_4}{\kappa_2}\theta^{\beta_2 - 1} -\tfrac{6\kappa_3^2}{\kappa_2}\theta^{2\beta_1 - 1} + \kappa_2\kappa_3 \left((m-1)\tfrac{\sqrt{2\pi}}{2} + \tfrac{1-2m}{8}\kappa_2 \sqrt{\theta}\right)\theta^{\beta_1} + o(\theta^{\bar{\beta}_\epsilon -1}).
\end{aligned}$$
\end{proof}

\section{Proofs and Verification in Section \ref{dist ver}}
\subsection{Proof of Lemma \ref{CHF condition}}
\label{proof of prop1}
\begin{proof}
We establish the convergence of the density approximation through Fourier methods. By direct verification and the definitions of $q^{(m, n)}_\theta(x)$ and $\varphi^{(m,n)}(x;\theta)$, we have the characteristic function approximation:
\begin{equation*}
\int_{-\infty}^\infty e^{\mathrm{i}ux} q^{(m,n)}_\theta(x) \, dx = \varphi^{(m,n)}(u;\theta) + o(\theta^{\bar{\beta}_\epsilon}).
\end{equation*}
By assumption \eqref{cond 2}, there exists some $v > 0$ such that $\varphi_X(u;\theta) \in L^1(\mathbb{R})$. Then the inverse Fourier transform ensures the existence of the density $p_\theta (x)$. Furthermore,
\begin{align*}
|p_\theta(x) - q^{(m,n)}_\theta(x)| 
&= \tfrac{e^{vx}}{2\pi} \left| \int_{-\infty}^\infty e^{-\mathrm{i}ux} \left(\varphi_X(u+\mathrm{i}v;\theta) - \varphi^{(m,n)}(u+\mathrm{i}v;\theta)\right) \mathrm{d}u \right| \\
&\leq \tfrac{e^{vx}}{2\pi} \int_{-\infty}^\infty \left|\varphi_X(u+\mathrm{i}v;\theta) - \varphi^{(m,n)}(u+\mathrm{i}v;\theta)\right| \mathrm{d}u.
\end{align*}

Given $\alpha > 0$ and $x_0 \in \mathbb{R}$, we have
\begin{align*}
&\sup_{x \leq x_0} (1 + x^2)^\alpha |p_\theta(x) - q^{(m,n)}_\theta(x)| \\
 \leq &\tfrac{1}{2\pi} \sup_{x \leq x_0} (1+x^2)^\alpha e^{vx} \int_{-\infty}^\infty \left|\varphi_X(u+\mathrm{i}v;\theta) - \varphi^{(m,n)}(u+\mathrm{i}v;\theta)\right| \mathrm{d}u \\
 =& o(\theta^{\bar{\beta}_\epsilon}),
\end{align*}
where the last equality follows from our assumption on $\varphi_X(\cdot + \mathrm{i}v; \theta)$ and the fact that $(1+x^2)^\alpha e^{vx}$ is bounded for $x \leq x_0$.
\end{proof}

\subsection{Proof of Proposition \ref{gamma}}
\label{proof gamma}

\begin{proof}
According to the model setup, the normalized random variable $X_\theta \equiv Z_\theta/\sigma_0(\theta)$ takes the characteristic function:
  \begin{equation*}
\varphi_X(u;\theta) = \exp\left(\tfrac{\mathrm{i} u w_\theta}{\sigma_0}\right)
\left(1 + \tfrac{\gamma_\theta^2 u^2}{\sigma_0^2}\right)^{-k_\theta}
\left(1 + \tfrac{ \mathrm{i} \gamma_\theta u}{\sigma_0}\right)^{-\bar{k}_\theta},
\end{equation*}
where $w_\theta = k_\theta \ln(1-\gamma_\theta) + (k_\theta + \bar{k}_\theta)\ln(1+\gamma_\theta)$ (note that $w_\theta$ is well defined since $\gamma_\theta \to 0$). The $n$-th cumulant of $X_\theta$ for $n \geq 2$ is:
\begin{equation}
\label{eq::kappas}
\tilde \kappa_n(\theta) = (n-1)! \tfrac{k_\theta + (-1)^n(k_\theta + \bar{k}_\theta)}{(2k_\theta + \bar{k}_\theta)^{n/2}}.
\end{equation}

Define $\beta_1 = \bar{\alpha} - \tfrac{3}{2}\alpha$ and $\beta_2 = -\alpha$, which yields the exponent:
$$
    \begin{aligned}\bar{\beta}_\epsilon &\equiv \min\{\beta_2 + \tfrac{1}{2}, 2\beta_1, 2\beta_2 - \epsilon, \beta_1 + \beta_2 - \epsilon\}\\
    &= \min\{-\alpha + \tfrac{1}{2}, -2\alpha -\epsilon, 2\bar{\alpha}-3\alpha, \bar{\alpha} -\tfrac{5}{2}\alpha - \epsilon\}.
\end{aligned}$$ 
We verify condition \eqref{cond 2} for $(m, n) = (0, 0)$ by showing that, for any $\epsilon > 0$,
\begin{equation*}
\int_{-\infty}^\infty |\varphi_X(u;\theta) - \varphi^{(0,0)}(u;\theta)|du = o(\theta^{\bar{\beta}_\epsilon}).
\end{equation*}

\noindent\textbf{Step 1: Expansion of the exponential term}

Note that the standard deviation $\sigma_0 = (2k_\theta + \bar k_\theta)^{1/2} \gamma_\theta$. Thus, by Taylor expansion of $\ln (x)$ around $x= 1$,
$$
\begin{aligned}
\exp\left(\tfrac{\mathrm{i}u w_\theta}{\sigma_0}\right) 
&= \exp\left(\tfrac{\mathrm{i}u}{\sigma_0}\left(k_\theta \ln(1-\gamma_\theta) + (k_\theta + \bar{k}_\theta)\ln(1+\gamma_\theta)\right)\right) \\
&= \exp\left(-\tfrac{\mathrm{i}u\sigma_0}{2} + \tfrac{\mathrm{i}u\bar{k}_\theta \gamma_\theta}{\sigma_0}\right) \cdot \exp\left(O\left(\tfrac{k_\theta \gamma_\theta^4}{\sigma_0}\right) + O\left(\tfrac{\bar{k}_\theta \gamma_\theta^3}{\sigma_0}\right)\right) \\
&= \exp\left(-\tfrac{\mathrm{i}u\sigma_0}{2} + \tfrac{\mathrm{i}u\bar{k}_\theta \gamma_\theta}{\sigma_0}\right) \cdot \left(1 + O\left(\theta^{\alpha + 4\beta - 1/2}\right) + O\left(\theta^{\bar{\alpha} + 3\beta - 1/2}\right)\right).
\end{aligned}
$$
Since $\alpha + 4\beta - \tfrac{1}{2} = \tfrac{3}{2} - \alpha > \bar{\beta}_\epsilon$ and $\bar{\alpha} + 3\beta - \tfrac{1}{2} = \bar \alpha - \tfrac{3}{2}\alpha + 1> \bar{\beta}_\epsilon$, we have:
\begin{equation*}
\exp\left(\tfrac{\mathrm{i}u w_\theta}{\sigma_0}\right) 
= \exp\left(-\tfrac{\mathrm{i}u\sigma_0}{2} + \tfrac{\mathrm{i}u\bar{k}_\theta \gamma_\theta}{\sigma_0}\right)
\left(1 + o(\theta^{\bar{\beta}_\epsilon})\right).
\end{equation*}

\noindent\textbf{Step 2: Simplification of characteristic function}

Substitute into $\varphi_X$:
$$
\begin{aligned}
\varphi_X(u;\theta) 
&= \exp\left(-\tfrac{\mathrm{i}u\sigma_0}{2} + \tfrac{\mathrm{i}u\bar{k}_\theta \gamma_\theta}{\sigma_0}\right)
\left(1 + \tfrac{\gamma_\theta^2 u^2}{\sigma_0^2}\right)^{-k_\theta}
\left(1 + \tfrac{\mathrm{i}\gamma_\theta u}{\sigma_0}\right)^{-\bar{k}_\theta} + o(\theta^{\bar{\beta}_\epsilon}) \\
&= \exp\left(-\tfrac{\mathrm{i}u\sigma_0}{2}\right)
\left(1 + \tfrac{\gamma_\theta^2 u^2}{\sigma_0^2}\right)^{-k_\theta - \bar{k}_\theta/2} \cdot \left(\tfrac{
\exp\left(\tfrac{2\mathrm{i}u\gamma_\theta}{\sigma_0}\right)
\left(1 + \tfrac{\gamma_\theta^2 u^2}{\sigma_0^2}\right)
}{\left(1 + \tfrac{\mathrm{i}u\gamma_\theta}{\sigma_0}\right)^2}\right)^{\bar{k}_\theta/2} 
+ o(\theta^{\bar{\beta}_\epsilon}).
\end{aligned}
$$

\noindent\textbf{Step 3: Asymptotic expansion of main terms}

Using the relationship $(1 + z/x)^{-x} = e^{-z}(1 + z^2/(2x) + O(x^{-2}))$,
$$
\begin{aligned}
\left(1 + \tfrac{\gamma_\theta^2 u^2}{\sigma_0^2}\right)^{-k_\theta - \bar{k}_\theta/2} 
&= e^{-u^2/2}\left(1 + \tfrac{u^4}{4(2k_\theta + \bar{k}_\theta)} + O\left(\theta^{-2\alpha}\right)\right) \\
&= e^{-u^2/2}\left(1 + \tfrac{\tilde{\kappa}_4}{24}u^4 + o(\theta^{\bar{\beta}_\epsilon})\right),
\end{aligned}
$$
where $\tilde \kappa_4 \equiv \tilde{\kappa}_4 (\theta) = \frac{6}{2k_\theta + \bar k_\theta}$, as given by \textnormal{Eq.} \eqref{eq::kappas}, is the excess kurtosis of $X_\theta$.
For the fractional term, we expand numerator and denominator under $|u| < \tfrac{\sigma_0}{\gamma_\theta}$ and obtain
$$\begin{aligned}
&\exp{\left(\tfrac{2\mathrm{i}u\gamma_\theta}{\sigma_0}\right)}\left(1 + \tfrac{\gamma_\theta^2u^2}{\sigma_0^2}\right) \\
=& 1 + \tfrac{2\mathrm{i}\gamma_\theta}{\sigma_0}u + \sum_{n=2}^\infty\left(\tfrac{1}{n!}\left(\tfrac{2\mathrm{i}\gamma_\theta}{\sigma_0}\right)^{n} + \tfrac{1}{(n-2)!}\tfrac{(2\mathrm{i})^{n-2}\gamma_\theta^n}{\sigma_0^n}\right)u^n,
\end{aligned}$$
and
$$\left(1 + \tfrac{\mathrm{i}\gamma_\theta}{\sigma_0}u\right)^{-2} = \sum_{n=0}^\infty (n+1)\left(\tfrac{-\mathrm{i}\gamma_\theta}{\sigma_0}\right)^nu^n.$$
By multiplying and collecting the coefficients of $u^n$, we have
$$\begin{aligned}\tfrac{\exp\left(\tfrac{2\mathrm{i} u\gamma_\theta}{\sigma_0}\right)\left(1 + \tfrac{\gamma_\theta^2 u ^2}{\sigma_0}\right)}{\left(1 
+ \tfrac{\mathrm{i}u\gamma_\theta}{\sigma_0^2}\right)^2} &= 1 + \tfrac{2\mathrm{i}\gamma_\theta^3}{3\sigma_0^3}u^3 + O\left(\tfrac{2\mathrm{i}\gamma_\theta^5}{\sigma_0^5}u^5\right)\\
&= 1+\tfrac{2\mathrm{i}\gamma_\theta^3}{3\sigma_0^3}u^3 + O(\theta^{5\beta - \tfrac{5}{2}}).
\end{aligned}$$
Raising to $\bar{k}_\theta/2$:
$$\begin{aligned}
\left(1 + \tfrac{2\mathrm{i}\gamma_\theta^3}{3\sigma_0^3}u^3 + O(\theta^{5\beta-5/2})\right)^{\bar{k}_\theta/2} 
&= 1 + \tfrac{\bar{k}_\theta}{2} \cdot \tfrac{2 \mathrm{i} \gamma_\theta^3}{3\sigma_0^3}u^3 + \tbinom{\bar{k}_\theta/2}{2} \left(\tfrac{2\mathrm{i}\gamma_\theta^3}{3\sigma_0^3}u^3\right)^2 + o(\theta^{\bar{\beta}_\epsilon}) \\
&= 1 + \tfrac{\mathrm{i} \bar{k}_\theta\gamma_\theta^3}{3\sigma_0^3}u^3 - \tfrac{\bar{k}_\theta^2\gamma_\theta^6}{18\sigma_0^6}u^6 + o(\theta^{\bar{\beta}_\epsilon})\\
&= 1 - \mathrm{i}\tfrac{\tilde \kappa_3}{6}u^3 - \tfrac{\tilde \kappa_3^2}{72}u^6 + o(\theta^{\bar{\beta}_\epsilon}),
\end{aligned}$$
where $\tilde \kappa_3 \equiv \tilde \kappa_3(\theta) = - \frac{2\bar k_\theta}{(2k_\theta + \bar k_\theta)^{3/2}}$ is the skewness of $X_\theta.$

\noindent\textbf{Step 4: Integral decomposition}

The error terms $o(\theta^{\bar{\beta}_\epsilon})$ above depend on a power function of $u$. Next, we extract the $u$ part and integrate the difference of $\varphi_X(u;\theta)$ and $ \varphi^{(0,0)}(u;\theta)$.

For $|u| < \sigma_0/\gamma_\theta$, we combine results and obtain
$$
\begin{aligned}
&|\varphi_X(u;\theta) - \varphi^{(0,0)}(u;\theta)| \\
=&e^{-u^2/2} \left|\left(1 - \mathrm{i}\tfrac{\tilde \kappa_3}{6}u^3 + \tfrac{\tilde\kappa_4}{24}u^4 - \tfrac{\tilde \kappa_3^2}{72}u^6 + P(u)o(\theta^{\bar \beta_\epsilon})\right) \right.\\
& \quad -\left. \left(1 -\mathrm{i}\kappa_3\theta^{\bar \alpha - \tfrac{3}{2}\alpha}u^3 + \kappa_4\theta^{-\alpha}u^4 - \tfrac{\kappa_3^2}{2} \theta^{2\bar\alpha - 3\alpha} u^6\right)\right|\\
=& e^{-u^2/2} P(u) o(\theta^{\bar{\beta}_\epsilon})
\end{aligned}
$$
uniformly for $|u| < \sigma_0/\gamma_\theta$, where $P(u)$ is some polynomial function and $o(\theta^{\bar{\beta}_\epsilon})$ is independent of $u$. The last equality holds due to:
$$\kappa_3(\theta) \theta^{\bar{\alpha} - \tfrac{3}{2}\alpha}= \tfrac{\tilde{\kappa}_3(\theta)}{6}, \quad \kappa_4(\theta)\theta^{-\alpha} = \tfrac{\tilde{\kappa}_4(\theta)}{24}.$$
Thus,
\begin{equation*}
\int_{|u| < \tfrac{\sigma_0}{\gamma_\theta}}|\varphi_X(u;\theta) - \varphi^{(0,0)}(u;\theta)|\mathrm{d}u
\lesssim o(\theta^{\bar{\beta}_\epsilon}) \int_{-\infty}^{\infty} e^{-u^2/2} P(u) \mathrm{d}u 
= o(\theta^{\bar{\beta}_\epsilon}).
\end{equation*}
For $|u| \geq \sigma_0/\gamma_\theta$, we integrate $|\varphi_X(u;\theta)|$ and substitute by $y = 1 + \gamma_\theta^2 u^2/\sigma_0^2$:
$$\begin{aligned}
\int_{|u| \geq \sigma_0/\gamma_\theta} |\varphi_X(u;\theta)| \mathrm{d}u 
&= \int_{|u| \geq \sigma_0/\gamma_\theta} \left(1 + \tfrac{\gamma_\theta^2 u^2}{\sigma_0^2}\right)^{-k_\theta - \bar{k}_\theta/2} \mathrm{d}u \\
&= \tfrac{\sigma_0}{\gamma_\theta} \int_2^\infty y^{-k_\theta - \bar{k}_\theta/2} \tfrac{\mathrm{d}y}{\sqrt{y-1}} \\
&\leq \tfrac{\sigma_0}{\gamma_\theta} \int_2^\infty y^{-k_\theta - \bar{k}_\theta/2} \mathrm{d}y \\
&= \tfrac{\sigma_0}{\gamma_\theta} \tfrac{2^{-k_\theta - \bar{k}_\theta/2 + 1}}{k_\theta + \bar{k}_\theta/2 - 1} \\
&= O\left(\theta^{-\alpha/2} e^{-c\theta^{\alpha}}\right) 
= o(\theta^{\bar{\beta}_\epsilon})
\end{aligned}$$
for some $c>0$. The same bound holds for $\varphi^{(0,0)}(u;\theta)$. Extension to complex number $u + \mathrm{i}v$ ($v>0$) follows directly since $\gamma_\theta/\sigma_0 \to 0$. The ATM asymptotics then follows from Theorem \ref{Theorem 2}-\ref{ATM skew}.
\end{proof}
\subsection{Proof of Proposition \ref{CGMY}}
\label{proof CGMY}
\begin{proof}
According to the model setup, the normalized random variable $X_\theta \equiv Z_\theta/\sigma_0(\theta)$ takes the characteristic function:
\begin{equation*}
\begin{aligned}
\varphi_X(u;\theta) &= \exp\left(\tfrac{\mathrm{i} uw_\theta}{\sigma_0} + C_\theta\Gamma(-Y)\left[\left(M_\theta -\tfrac{\mathrm{i} u}{\sigma_0}\right)^Y - M_\theta^Y + \left(G_\theta + \tfrac{\mathrm{i} u}{\sigma_0}\right)^Y-G_\theta^Y\right]\right),
\end{aligned}
\end{equation*}
where $w_\theta = -C_\theta \Gamma(-Y)\left[(M_\theta - 1)^Y - M_\theta^Y + (G_\theta+1)^Y - G_\theta^Y\right]$. The $n$-th cumulant of $X_\theta$ is:
\begin{equation}
\label{eq::kappas2}
\tilde \kappa_n(\theta) = \tfrac{C_\theta \Gamma(n-Y)\left(M_\theta^{Y-n} + (-1)^n G_\theta^{Y-n}\right)}{\sigma_0^n}.
\end{equation}

Define $\beta_1 = \alpha_G - 2\alpha_M - \tfrac{1}{2} >0$ and $\beta_2 = -2\alpha_M -1 > 0$, which yields the exponent:
$$
    \begin{aligned}\bar{\beta}_\epsilon &\equiv \min\{\beta_2 + \tfrac{1}{2}, 2\beta_1, 2\beta_2 - \epsilon, \beta_1 + \beta_2 - \epsilon\}\\
    &= \min\{-2\alpha_M -1/2, 2\alpha_G-4\alpha_M-1, -4\alpha_M-2-\epsilon, \alpha_G-4\alpha_M-3/2 - \epsilon\}.
\end{aligned}$$ 
We verify condition \eqref{cond 2} for $(m, n) = (0, 0)$ by first showing:
\begin{equation*}
\int_{-\infty}^\infty |\varphi_X(u;\theta) - \varphi^{(0,0)}(u;\theta)|du = o(\theta^{\bar{\beta}_\epsilon}).
\end{equation*}

\noindent\textbf{Step 1: Expansion of the exponential term}

We begin with Taylor expansion of the term $\exp\left(\tfrac{\mathrm{i}u w_\theta}{\sigma_0}\right)$:
\begin{align*}
\exp\left(\tfrac{\mathrm{i}u w_\theta}{\sigma_0}\right) 
&= \exp\left(\tfrac{\mathrm{i} u}{\sigma_0}C_\theta \Gamma(1-Y)\left(G_\theta^{Y-1} - M_\theta^{Y-1}\right)- \tfrac{iu\sigma_0}{2} + o(\theta^{\bar{\beta}_\epsilon})\right).
\end{align*}
For the second exponent, we note that the variance satisfies
$$\sigma_0^2 = C_\theta \Gamma(2-Y) (M_\theta^{Y-2} + G_\theta^{Y-2}).$$
Thus, expanding the power functions around $x = 1$ yields:
\begin{align*}
&C_\theta\Gamma(-Y)\left(M_\theta^Y\left(\left(1 -\tfrac{\mathrm{i} u}{M_\theta\sigma_0}\right)^Y - 1\right)  + G_\theta^Y\left(\left(1 + \tfrac{\mathrm{i} u}{G_\theta\sigma_0}\right)^Y-1\right)\right) \\
=& \tfrac{\mathrm{i} u}{\sigma_0}C_\theta \Gamma(1-Y)\left(M_\theta^{Y-1} - G_\theta^{Y-1}\right) - \tfrac{u^2}{2} - \tfrac{\mathrm{i} u^3}{6\sigma_0^3}C_\theta\Gamma(3-Y)\left(M_\theta^{Y-3} - G_\theta^{Y-3}\right) \\
&\quad + \tfrac{u^4}{24\sigma_0^4}C_\theta\Gamma(4-Y)\left(M_\theta^{Y-4} + G_\theta^{Y-4}\right) + o(\theta^{\bar{\beta}_\epsilon}),
\end{align*}
where the error term comes from the combined asymptotics of the odd-order term, $n = 5$, $O(\theta^{\alpha_G - 4\alpha_M -\tfrac{3}{2}})= o(\theta^{\bar{\beta}_\epsilon})$ and the even-order term, $n = 6$,  $O(\theta^{\alpha_M(Y-6) -3 + \beta}) = o(\theta^{\bar{\beta}_\epsilon})$ due to $\alpha_M(Y - 6) + 3 + \beta = -4\alpha_M - 2 > \bar \beta_{\epsilon}$.

\noindent\textbf{Step 2: Simplification of characteristic function}

Substitute expansions into $\varphi_X$:
\begin{align*}
\varphi_X(u;\theta) 
&= \exp\left(\tfrac{\mathrm{i} u}{\sigma_0}C_\theta \Gamma(1-Y)\left(G_\theta^{Y-1} - M_\theta^{Y-1}\right)\right) \cdot \exp\left(-\tfrac{\mathrm{i} u\sigma_0}{2} - \tfrac{u^2}{2}\right) \\
&\quad \cdot \exp\left(\tfrac{\mathrm{i} u}{\sigma_0}C_\theta \Gamma(1-Y)(M_\theta^{Y-1} - G_\theta^{Y-1}) -\tfrac{\mathrm{i} u^3}{6\sigma_0^3}C_\theta\Gamma(3-Y)(M_\theta^{Y-3} - G_\theta^{Y-3})  \right) \\
&\quad \cdot \exp\left(\tfrac{u^4}{24\sigma_0^4}C_\theta\Gamma(4-Y)\left(M_\theta^{Y-4} + G_\theta^{Y-4}\right)+ o(\theta^{\bar{\beta}_\epsilon})\right).
\end{align*}
The first-order term cancels out, yielding:
\begin{align*}
\varphi_X(u;\theta) 
&= \exp\left(-\tfrac{\mathrm{i}u\sigma_0}{2} - \tfrac{u^2}{2}\right) \cdot \left(1 - \tfrac{\mathrm{i}u^3}{6\sigma_0^3}C_\theta\Gamma(3-Y)\left(M_\theta^{Y-3} - G_\theta^{Y-3}\right)\right. \\
&\quad + \tfrac{u^4}{24\sigma_0^4}C_\theta\Gamma(4-Y)\left(M_\theta^{Y-4} + G_\theta^{Y-4}\right) \\
&\quad \left.- \tfrac{u^6}{72\sigma_0^6}\left[C_\theta\Gamma(3-Y)\left(M_\theta^{Y-3} - G_\theta^{Y-3}\right)\right]^2+ o(\theta^{\bar{\beta}_\epsilon})\right)\\
&= \exp\left(-\tfrac{\mathrm{i} u\sigma_0}{2} - \tfrac{u^2}{2}\right) \left(1 - \tfrac{\mathrm{i}}{6}\tilde{\kappa}_3u^3 + \tfrac{1}{24}\tilde{\kappa}_4u^4- \tfrac{1}{72}\tilde{\kappa}_3^2u^6 + o(\theta^{\bar{\beta}_\epsilon}) \right),
\end{align*}
where $\tilde \kappa_3 \equiv \tilde{\kappa}_3(\theta)$ and $\tilde{\kappa}_4(\theta)$, given by \eqref{eq::kappas2}, are the skewness and excess kurtosis of $X_\theta$, respectively.

\noindent\textbf{Step 3: Integral decomposition}

The error terms $o(\theta^{\bar{\beta}_\epsilon})$ above depend on a power function of $u$. Next, we extract the $u$ part and integrate the difference of $\varphi_X(u;\theta)$ and $ \varphi^{(0,0)}(u;\theta)$.

By combining the results above, the difference takes the order:
\begin{align*}
&|\varphi_X(u;\theta) - \varphi^{(0,0)}(u;\theta)| \\
=& e^{-u^2/2}\left|\left(1 - \tfrac{\mathrm{i}}{6}\tilde{\kappa}_3u^3 + \tfrac{1}{24}\tilde{\kappa}_4u^4 - \tfrac{1}{72}\tilde{\kappa}_3^2u^6 + P(u)o(\theta^{\bar \beta_\epsilon})\right) \right.\\
&\quad - \left.\left(1 - \mathrm{i}\kappa_3\theta^{\alpha_G-2\alpha_M-1/2}u^3 + \kappa_4\theta^{-2\alpha_M-1}u^4 - \tfrac{1}{2}\kappa_3^2\theta^{2(\alpha_G-2\alpha_M-1/2)}u^6\right)\right| \\
=& e^{-u^2/2} P(u) o(\theta^{\bar{\beta}_\epsilon}),
\end{align*}
uniformly for $|u| < M_\theta\sigma_0$, where $P(u)$ is a polynomial function, $o(\theta^{\bar{\beta}_\epsilon})$ is independent of $u$, and the last equality holds due to 
\begin{equation*}
\kappa_3(\theta)\theta^{\alpha_G - 2\alpha_M -1/2} = \tfrac{\tilde{\kappa}_3(\theta)}{6}, \quad
\kappa_4(\theta)\theta^{-2\alpha_M-1} = \tfrac{\tilde{\kappa}_4(\theta)}{24}.
\end{equation*} 
Thus,
\begin{equation*}
\int_{|u| < M_\theta \sigma_0}|\varphi_X(u;\theta) - \varphi^{(0,0)}(u;\theta)|\mathrm{d}u
\lesssim o(\theta^{\bar{\beta}_\epsilon}) \int_{-\infty}^{\infty} e^{-u^2/2} P(u) du 
= o(\theta^{\bar{\beta}_\epsilon}).
\end{equation*}\
And for the tail region $|u| \geq M_\theta\sigma_0$, we note that
$$\begin{aligned}\operatorname{Re}\left(1 - \tfrac{\mathrm{i}u}{M_\theta\sigma_0}\right)^Y &= \left(1 + \tfrac{u^2}{M_\theta^2\sigma_0^2 }\right)^{Y/ 2}\cos(Y \arctan \tfrac{u}{M_\theta \sigma_0})\\
&\sim \tfrac{|u|^Y}{(M(\theta)\sigma_0)^Y}\cos{(\tfrac{\pi}{2}Y)},\quad \text{as } |u| \to \infty,
\end{aligned}$$
which results in
$$\begin{aligned}
   & \left|\exp\left(C_\theta \Gamma(- Y)\left((M_\theta - \tfrac{\mathrm{i}u}{\sigma_0})^Y - M_\theta^Y\right)\right)\right|\\
   = &\exp\left(\Gamma(- Y)C_\theta  M_\theta^Y\left(\operatorname{Re}(1 -\tfrac{\mathrm{i}u}{\sigma_0 M_\theta})^Y - 1\right)\right)\\
   \sim & \exp\left(\Gamma(- Y)C_\theta \tfrac{|u|^Y}{\sigma_0^Y}\cos{(\tfrac{\pi}{2}Y)}\right)\\
   \le & \exp(-c \theta^b |u|^Y)
\end{aligned}$$
for $b := (1 + 2\alpha_M)(1 - Y / 2) < 0$ and some constant $c > 0$. The same argument can be applied to $\operatorname{Re}\left(1 + \tfrac{\mathrm{i}u}{\sigma_0 G_\theta}\right)^Y$. Thus, we also obtain
$$\left|\exp\left(C_\theta \Gamma(- Y)\left((G_\theta + \tfrac{\mathrm{i}u}{\sigma_0})^Y - G_\theta^Y\right)\right)\right| \lesssim \exp(-c \theta^b |u|^Y),$$
where $\lesssim$ means the ratio of left and right side expressions is asymptotically less or equal to $1.$ It then follows that
\begin{align*}
&\int_{|u| \geq M_\theta\sigma_0} |\varphi_X(u;\theta)| \mathrm{d}u \\
\lesssim &  \int_{|u| \geq M_\theta\sigma_0} \exp(-2c \theta^b |u|^Y)  \mathrm{d}u \\
\le & 2 \tfrac{(M_\theta \sigma_0)^{1 - Y}}{c\theta^b} \exp(-2c\theta^b (M_\theta \sigma_0)^Y) \\
=& O\left(\exp\left(-\tilde c\theta^{1-2\alpha_M}\right)\right) = o(\theta^{\bar{\beta}_\epsilon})
\end{align*}
for some constant $\tilde c > 0$. 

Moreover, by direct computation, 
$$\int_{|u| \geq M_\theta\sigma_0} |\varphi_X(u;\theta)| = o(\theta^{\bar{\beta}_\epsilon}).$$
Therefore, combing the results above, condition \eqref{cond 2} holds true for $v = 0$. Extension of variable from $u \in \mathbb{R}$ to $u + \mathrm{i}v$ ($v>0$) follows directly from $M_\theta \sigma_0(\theta) \to \infty$. And the ATM asymptotics results follow from Theorem \ref{Theorem 2}-\ref{ATM skew}.
\end{proof}

\subsection{Verification of the Condition for Kellerer's Theorem}
\label{cond Kellerer}
Kellerer's theorem (cf. \cite{kellerer1972markov}) states: Let $\{p_t^Z(x)\}_{t\in (0, 1]}$ be a family of marginal densities of $\{e^{Z_t}\}_{t\in (0,1]}$, with finite and constant expectations, such that for $s \le t$ the density at time $t$ dominates the density at time $s$ in convex order, i.e., formula \eqref{convex order}. Then there exists a Markov martingale with marginal densities of $\{p_t(x)\}_{t \in (0,1]}$.

The convex-order condition \eqref{convex order} is equivalent to (cf. Theorem 3.A.1, \cite{shaked2007stochastic}), for $s \le t$,
\begin{equation}
\label{convex order 2}
E[(e^{Z_s} - k)_+] \le E[(e^{Z_t} - k)_+], \quad \forall k \in \mathbb{R},
\end{equation}
where $E[(e^{Z_t} - k)_+]$ is a call function.

\begin{figure}[!htbp]
    \centering
    \includegraphics[width=\linewidth]{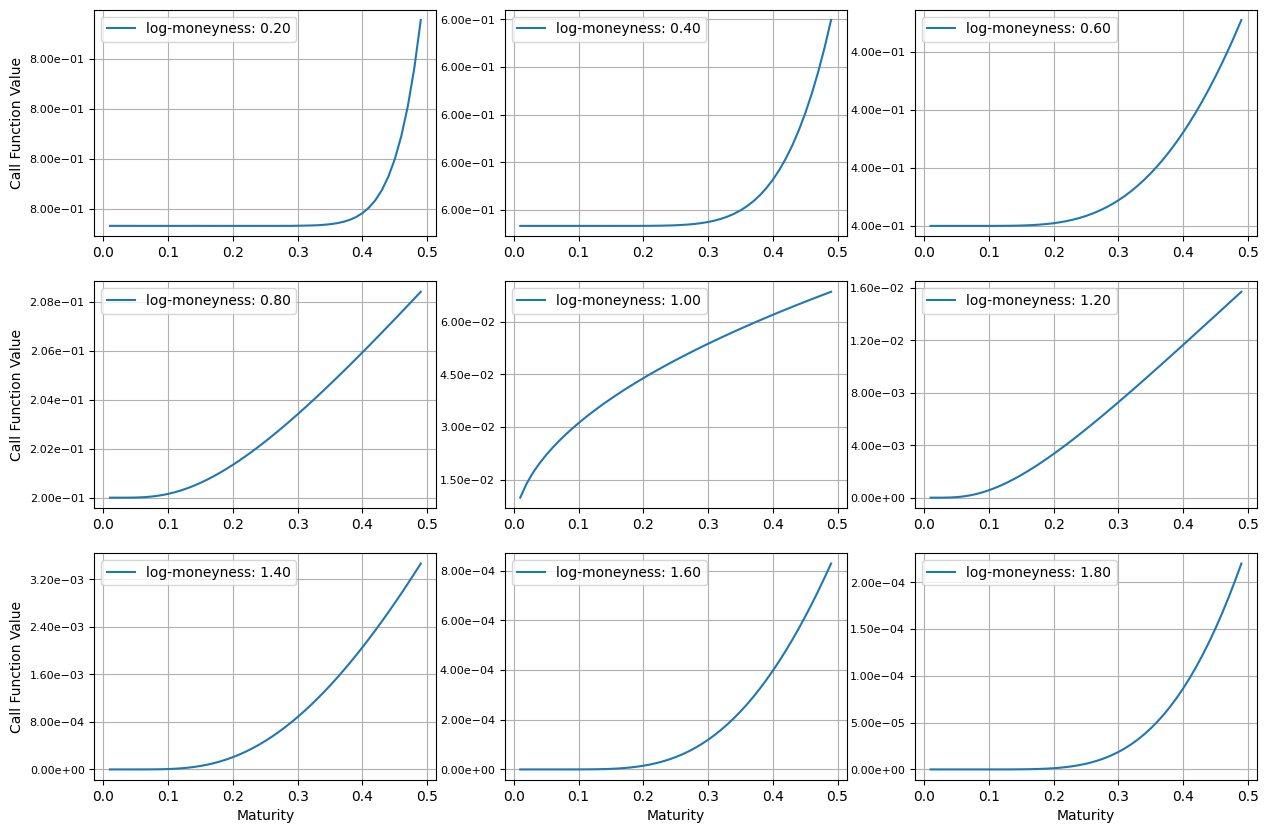}
    \caption{Validation of convex-order condition for the scalable gamma martingale model with parameters $(\alpha, \bar{\alpha}) = (-0.2, -0.1)$.}
    \label{gamma_kellerer}
\end{figure}

\begin{figure}[!htbp]
    \centering
    \includegraphics[width=\linewidth]{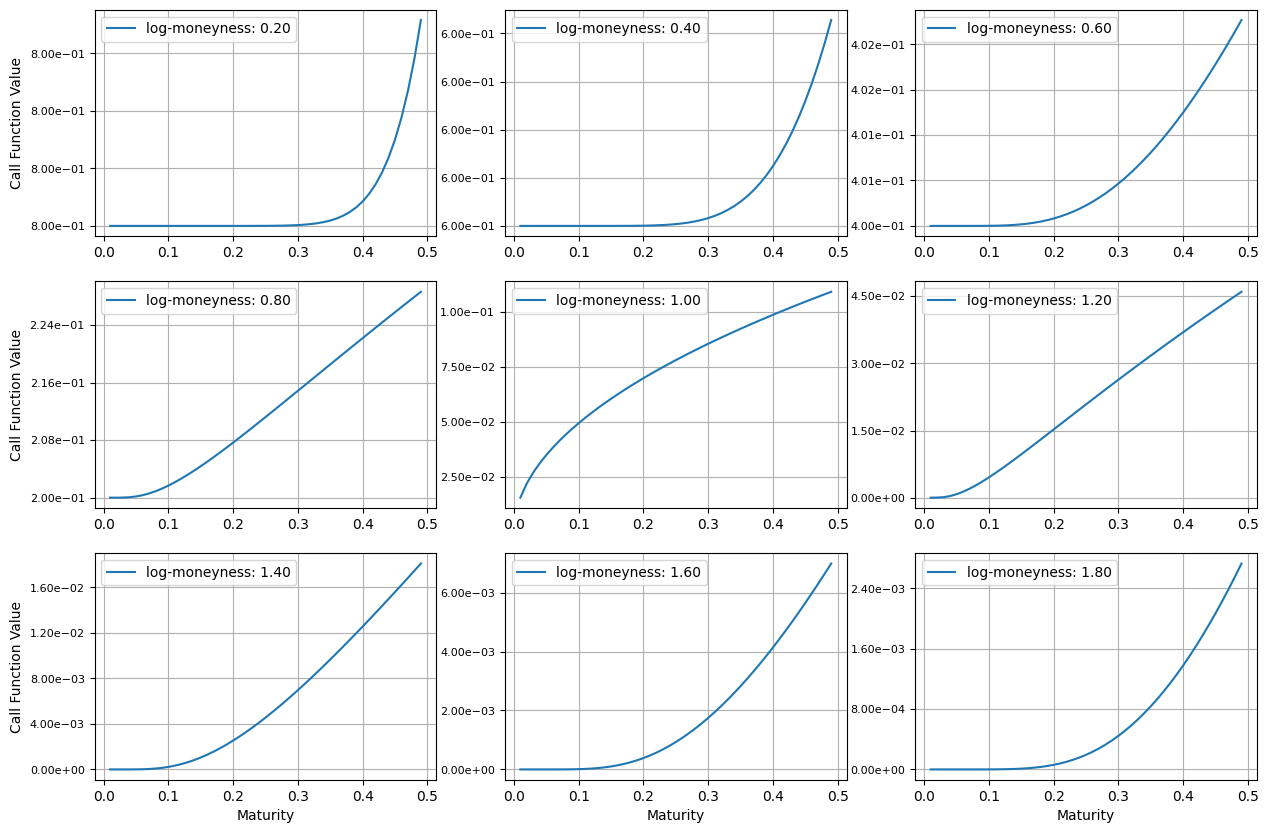}
    \caption{Validation of convex-order condition for the scalable CGMY martingale model with parameters $(\alpha_M, \alpha_G) = (-0.6, -0.5)$.}
    \label{cgmy_kellerer}
\end{figure}

In the following, we numerically validate that the implementations of distribution-based models in Figure \ref{num_exam} satisfy condition \eqref{convex order 2}. The verification is shown in Figure \ref{gamma_kellerer} and \ref{cgmy_kellerer}, with $(\alpha, \bar{\alpha}) = (-0.2, -0.1)$ and $(\alpha_M, \alpha_G) = (-0.6, -0.5)$, respectively. We examine the condition for maturities $\theta \in [0.01, 0.5]$ and log-moneyness $k \in [0.2, 1.8]$, with scalable Gamma model (Figure \ref{gamma_kellerer}) and scalable CGMY model (Figure \ref{cgmy_kellerer}). 

The experiments show that condition \eqref{convex order 2} holds for all the six pairs of parameterization considered in Figure \ref{num_exam} since all call function values are monotonically increasing, which then ensures the existence of a martingale that aligns with the densities of $S_t = S_0 e^{Z_t}$. 

\section{Proofs and Supplements in Section \ref{svm ver}}
\subsection{Proof of Proposition \ref{substitute}}
\label{proof of prop2}
\begin{proof}
We first establish the relationship between $\sigma_0^2$ and $\tilde{\sigma}_0^2$. By the definitions of $\sigma_0$ and $\tilde \sigma_0$, we obtain:
\begin{align*}
|\sigma_0^2 - \tilde{\sigma}_0^2| 
&= \left|\operatorname{Var}(Z_\theta) - \left(\int_0^\theta E[v_t] \mathrm{d}t\right)^2\right| \\
&= \left|\tfrac{1}{4}\operatorname{Var}\left(\int_0^\theta v_t \mathrm{d}t\right) - E\left[\int_0^\theta \sqrt{v_t}\mathrm{d}B_t \cdot \int_0^\theta v_t \mathrm{d}t\right]\right| \\
& \leq \tfrac{1}{4}\operatorname{Var}\left(\int_0^\theta v_t \mathrm{d}t\right) +E\left[\left|\int_0^\theta \sqrt{v_t}\mathrm{d}B_t\right|  \left|\int_0^\theta (v_t-E[v_t]) \mathrm{d}t\right|\right]\\
&\leq \tfrac{1}{4}\operatorname{Var}\left(\int_0^\theta v_t \mathrm{d}t\right) + \tilde{\sigma}_0\sqrt{\operatorname{Var}\left(\int_0^\theta v_t \mathrm{d}t\right)}.
\end{align*}
Rewriting in terms of the time-averaged variance yields:
\begin{align*}
|\sigma_0^2 - \tilde{\sigma}_0^2| &\le \tfrac{1}{4}\theta^2\operatorname{Var}\left(\tfrac{1}{\theta}\int_0^\theta v_t \mathrm{d}t\right) + \theta\tilde{\sigma}_0\sqrt{\operatorname{Var}\left(\tfrac{1}{\theta}\int_0^\theta v_t \mathrm{d}t\right)} \\
&= o(\theta^{3/2}),
\end{align*}
where the last equality follows from the assumption that $\lim_{\theta \downarrow 0}\operatorname{Var}\left(\frac{1}{\theta}\int_0^\theta v_t \mathrm{d}t\right) = 0$. This implies the following bound on the difference of standard deviations:
\begin{equation*}
|\sigma_0 - \tilde{\sigma}_0| = \tfrac{|\sigma_0^2 - \tilde{\sigma}_0^2|}{\sigma_0 + \tilde{\sigma}_0} = o(\theta).
\end{equation*}

Using the above result, we expand the density functions and the Hermite polynomials $H_n$ around $\tilde{\sigma}_0$:
    $$
    \begin{aligned}\phi(x + \tfrac{\sigma_0}{2}) &= \phi(x+\tfrac{\tilde{\sigma}_0}{2}) + o(\theta),\\
     H_n(x + \tfrac{\sigma_0}{2}) &= H_n(x+ \tfrac{\tilde{\sigma}_0}{2}) + o(\theta),\quad n \ge 1,\end{aligned}$$
from which, and the definitions of $q^{(1,0)}_\theta$ and $\tilde q_\theta$, $|q^{(1,0)}_\theta(x) - \tilde{q}_\theta(x)| = o(\theta^{2H}).$ Since both $q^{(1,0)}_\theta(x)$ and $\tilde{q}_\theta(x)$ belong to the Schwartz space $\mathcal{S}(\mathbb{R})$, their difference $
q^{(1,0)}_\theta(x) - \tilde{q}_\theta(x)$ is also Schwartz, which then ensures condition \eqref{density convergence} for any $\alpha > 0$. This completes the proof.
\end{proof}
\subsection{Proof of Proposition \ref{decaying leverage}}
\label{proof decaying leverage}
Since this is a non-trivial extension of the results established in Section 4 of \cite{euch2019short}, we develop our proofs from the start while selectively applying results from their work when they can be directly used without ambiguity.

\begin{lemma}
\label{lemma 1}
    Under the model setup of Proposition \ref{decaying leverage}, we denote $M_\theta = \int_0^\theta \sqrt{v_t}\mathrm{d}B_t$, then there exists a family of random vectors $$\{(M_\theta^{(0)}, M_\theta^{(1)}, M_\theta^{(2)}, M_\theta^{(3)});\quad \theta\in (0,1)\}$$
    such that
    \begin{itemize}
        \item $M_\theta^{(0)}$ follows a standard normal distribution for all $\theta > 0$,
        \item $$\sup_{\theta\in (0,1)}\|M_\theta^{(i)}\|_p < \infty,\quad i=1,2,3 $$
        for all $p > 0$, 
        \item for any $\epsilon \in (0, \tfrac{1}{8})$,
        \begin{equation}
        \label{epsilon cond}\begin{aligned}
            & \lim_{\theta\to 0} \theta^{-\bar{\beta}-2\epsilon}\left\|\tfrac{M_\theta}{\tilde{\sigma}_0} - M_\theta^{(0)} - \theta^{1/2}M_\theta^{(1)} -\theta M_\theta^{(2)}\right\|_{1 + \epsilon} = 0,\\
              & \lim_{\theta\to 0} \theta^{1/2-\bar{\beta}-2\epsilon}\left\|\tfrac{\langle M\rangle_\theta}{\tilde{\sigma}_0^2} - 1 - \theta^{1/2}M_\theta^{(3)}\right\|_{1 + \epsilon} = 0,
        \end{aligned}\end{equation}
        where $\bar{\beta} = \min\{1 + 2\alpha_\rho, 3/2\}$, and
        \item the derivatives
        \begin{equation}
        \label{abc}\begin{aligned}
& a_\theta^{(i)}(x)=\tfrac{\mathrm{d}}{\mathrm{~d} x}\left(E\left[M_\theta^{(i)} \mid M_\theta^{(0)}=x\right] \phi(x)\right), \quad i=1,2,3, \\
& b_\theta(x)=\tfrac{\mathrm{d}^2}{\mathrm{~d} x^2}\left(E\left[M_\theta^{(1)} M_\theta^{(3)}\mid M_\theta^{(0)}=x\right] \phi(x)\right), \\
& c_\theta(x)=\tfrac{\mathrm{d}^2}{\mathrm{~d} x^2}\left(E\left[\left|M_\theta^{(1)}\right|^2 \mid M_\theta^{(0)}=x\right] \phi(x)\right)
\end{aligned}\end{equation}
exist in Schwartz space.
    \end{itemize}
\end{lemma}
\begin{proof}
    We recall that $f = \sqrt{v}$, $g = f^\prime c$ and $h = v^\prime c$, and we denote by $\mathcal{L}$ the infinitesimal generator of diffusion process $Y$. By It\^{o}'s formula, 
    $$
    \begin{aligned}
M_\theta & =f\left(Y_0\right) B_\theta+\int_0^\theta \int_0^t g\left(Y_s\right) \mathrm{d} W_s \mathrm{~d} B_t+\int_0^\theta \int_0^t \mathcal{L} f\left(Y_s\right) \mathrm{d} s \mathrm{~d} B_t,\\
\langle M\rangle_\theta & =v\left(Y_0\right) \theta+\int_0^\theta \int_0^t h\left(Y_s\right) \mathrm{d} W_s \mathrm{~d} t+\int_0^\theta \int_0^t \mathcal{L} v\left(Y_s\right) \mathrm{d} s \mathrm{~d} t.
\end{aligned}
$$
Let $\bar{B}_t^\theta=\theta^{-1 / 2} B_{\theta t}$, $\bar{W}_t^\theta=\theta^{-1 / 2} W_{\theta t}$, and $Y_t^\theta=Y_{\theta t}$. Then
$$
\begin{aligned}
\tfrac{M_\theta}{\sqrt{\theta}} & =f\left(Y_0\right) \bar{B}_1^\theta+\sqrt{\theta} \int_0^1 \int_0^u g\left(Y_v^\theta\right) \mathrm{d} \bar{W}_v^\theta \mathrm{d} \bar{B}_u^\theta+\theta \int_0^1 \int_0^u \mathcal{L} f\left(Y_v^\theta\right) \mathrm{d} v \mathrm{~d} \bar{B}_u^\theta, \\
\tfrac{\langle M\rangle_\theta}{\theta} & =v\left(Y_0\right)+\sqrt{\theta} \int_0^1 \int_0^u h\left(Y_v^\theta\right) \mathrm{d} \bar{W}_v^\theta \mathrm{d} u+\theta \int_0^1 \int_0^u \mathcal{L} v\left(Y_v^\theta\right) \mathrm{d} v \mathrm{~d} u.
\end{aligned}$$
Since, for $v \in (0,1)$, $E[\mathcal{L}v(Y^\theta_v)- \mathcal{L}v(Y_0)] = O(\theta),$ we have
\[
\tfrac{\tilde{\sigma}_0^2}{\theta} = \tfrac{E[\langle M\rangle_{\theta}]}{\theta} = v(Y_0) + \tfrac{1}{2} \mathcal{L}v(Y_0)\theta + O(\theta^{2}),
\]
and thus
\[
\tfrac{\tilde{\sigma}_0}{\sqrt{\theta}} = f(Y_0) + \tfrac{1}{4} \tfrac{\mathcal{L}v(Y_0)}{f(Y_0)} \theta + O(\theta^{2}).
\]
We let $M_\theta^{(0)} = \bar{B}_1^\theta$ and 
\[
M_{\theta}^{(1)} = \tfrac{g(Y_0)}{f(Y_0)} \int_0^1 \bar{W}_u^\theta \mathrm{d}\bar{B}_u^\theta,
\]

\[
M_{\theta}^{(2)} = -\tfrac{\mathcal{L}v(Y_0)}{4v(Y_0)} \bar{B}_1^\theta + \tfrac{g'(Y_0)c(Y_0)}{f(Y_0)} \int_0^1 \int_0^u \bar{W}_v^\theta \mathrm{d}\bar{W}_v^\theta \mathrm{d}\bar{B}_u^\theta + \tfrac{\mathcal{L}f(Y_0)}{f(Y_0)} \int_0^1 u \mathrm{d}\bar{B}_u^\theta.
\]
By expanding the division $\tfrac{M_\theta}{\tilde{\sigma}_0} = \tfrac{M_\theta}{\sqrt{\theta}} / \tfrac{\tilde \sigma_0}{\sqrt{\theta}}$, we have
$$\begin{aligned}
    \tfrac{M_\theta}{\tilde\sigma_0} &= \tfrac{M_\theta}{\sqrt{\theta}} \tfrac{1}{f(Y_0) (1 + \tfrac{1}{4}\tfrac{\mathcal{L}v(Y_0)}{f^2(Y_0)}\theta)} + O(\theta^2)\\
    &= \tfrac{M_\theta}{\sqrt{\theta}f(Y_0)}(1 - \tfrac{1}{4}\tfrac{\mathcal{L}v(Y_0)}{f^2(Y_0)}\theta) + O(\theta^2)\\
    &=\left(\bar{B}_1^\theta + \tfrac{\sqrt{\theta}}{f(Y_0)}\int_0^1 \int_0^u \left(g(Y_0) + \sqrt{\theta}g'c(Y_0)\bar W_v^\theta + \theta \mathcal{L}g(Y_0) v\right) \mathrm{d} \bar{W}_v^\theta \mathrm{d} \bar{B}_u^\theta\right)\\
    & \quad \quad + \left(\tfrac{\theta}{f(Y_0)}\int_0^1 \int_0^u \left(\mathcal{L} f\left(Y_0\right) + \sqrt{\theta}(\mathcal{L}f)'c(Y_0) \bar W_v^\theta \right) \mathrm{d} v \mathrm{~d} \bar{B}_u^\theta\right)\\
    & \quad \quad - \left(\theta \tfrac{\mathcal{L}v(Y_0)}{4v(Y_0)} \bar{B}_1^\theta + \theta^{3/2}\tfrac{1}{4}\tfrac{\mathcal{L}v(Y_0)g(Y_0)}{f^3(Y_0)}\int_0^1 \bar W_u^\theta \mathrm{d}\bar B_u^\theta\right)  +  O(\theta^2)\\
    &= M_\theta^{(0)} + \sqrt{\theta}M_\theta^{(1)} + \theta M_\theta^{(2)} + \theta^{3/2}\tfrac{\mathcal{L}g(Y_0)}{f(Y_0)}\int_0^1 \int_0^u v \mathrm{~d} \bar{W}_v^\theta\mathrm{~d} \bar{B}_u^\theta\\
    & \quad\quad + \theta^{3/2}\left(\tfrac{(\mathcal{L}f)'c(Y_0)}{f(Y_0)}\int_0^1 \int_0^u  \bar W_v^\theta \mathrm{~d}v\mathrm{~d}\bar B_u^\theta - \tfrac{\mathcal{L}v(Y_0)g(Y_0)}{4f^3(Y_0)}\int_0^1 \bar W_u^\theta \mathrm{~d}\bar B_u^\theta\right) + O(\theta^2).
\end{aligned}$$
Since $\mathrm{d}\langle \bar W_t^\theta, \bar B_t^\theta\rangle = O(\theta^{\alpha_\rho})\mathrm{d}t$ by assumption, we obtain, e.g., from Burkholder-Davis-Gundy inequality, that for any $\epsilon > 0$,
$$\left\|\tfrac{M_\theta}{\tilde{\sigma}_0} -M_\theta^{(0)} - \sqrt{\theta}M_\theta^{(1)} - \theta M_\theta^{(2)}\right\|_{1 + \epsilon} = O(\theta^2) + O(\theta^{\tfrac{3}{2} + \alpha_\rho}).$$
Following a similar argument, by denoting \[
M_{\theta}^{(3)} = 2 \tfrac{g(Y_0)}{f(Y_0)} \int_0^1 \bar{W}_u^\theta \, du,
\]
we have
$$\left\|\tfrac{\langle M\rangle_\theta}{\tilde{\sigma}_0^2} - 1 - \sqrt{\theta}M_\theta^{(3)} \right\|_{1+\epsilon} = O(\theta^{\tfrac{3}{2}}) + O(\theta^{1 + \alpha_\rho}).$$

Since $\min\{2, \tfrac{3}{2} + \alpha_\rho\} - \bar{\beta} - 2\epsilon > 0$ and $\min\{\tfrac{3}{2}, 1 + \alpha_\rho\} + \tfrac{1}{2} - \bar{\beta} -2\epsilon > 0$ hold for $0 < \epsilon < \tfrac{1}{8}$, we have shown that \textnormal{Eq.} \eqref{epsilon cond} is satisfied. Moreover, we obtain from the explicit expressions in \cite{euch2019short} that $a_\theta^{(i)}(x), \, i= 1,2,3$ and $c_\theta(x)$ exist with
\begin{equation}
\label{abc val}\begin{aligned}
&a^{(1)}_\theta(x) = -\kappa_3(\theta)\theta^{\alpha_\rho}H_3(x)\phi(x)\\
&a^{(3)}_\theta(x) = -2 \kappa_3(\theta) \theta^{\alpha_\rho}H_2(x)\phi(x)\\
& a^{(2)}_\theta(x) - \tfrac{1}{2}c_\theta(x) = -\kappa_4(\theta)H_4(x)\phi(x) - \tfrac{\kappa_3(\theta)^2}{2} H_6(x)\phi(x),
\end{aligned}\end{equation}
with \begin{equation}
\label{kappa34}
\kappa_3(\theta) = \tfrac{\rho g}{2f}(Y_0), \quad\kappa_4(\theta) = \tfrac{\rho_\theta^2}{6}\tfrac{g^\prime c}{f}(Y_0) + \tfrac{1 + 2\rho_\theta^2}{6}\tfrac{g^2}{f^2}(Y_0) = \tfrac{1}{6}\tfrac{g^2}{f^2}(Y_0)  + O(\theta^{2\alpha_\rho}),\end{equation}
and that each $a_\theta^{(i)}$, $i = 1, 2, 3$ and $c_\theta(x)$ are in Schwartz space. Moreover, we apply the results on the conditional expectations of Wiener-It\^{o} integrals (see \cite{nualart1988moments}) and show that
 \begin{equation}
    \label{b}\begin{aligned}
     & b_\theta(x) = 4\kappa_4(\theta)H_3(x)\phi(x) + O(\theta^{2\alpha_\rho}),
 \end{aligned}
 \end{equation}
 with $b_\theta(x)$ in Schwartz space. Thus, \textnormal{Eq.} \eqref{abc} is satisfied and the results of this proposition follow.
\end{proof}

Next, we define
$$\tilde X_\theta = M_\theta^{(0)} - \tfrac{\tilde{\sigma}_0}{2} + \sqrt{\theta}M_\theta^{(1)} + \theta M_\theta^{(2)} - \tfrac{\tilde{\sigma}_0}{2} \sqrt{\theta}M_\theta^{(3)}.$$
\begin{lemma}
\label{int diff}
    For any $\epsilon \in (0,\tfrac{1}{8})$ and $\alpha \in \mathbb{N} \cup \{0\}$,
    $$\sup_{|u| \le \theta^{-\epsilon}}\left| E[X_\theta^\alpha e^{\mathrm{i}uX_\theta}] - E[\tilde X_\theta^\alpha e^{\mathrm{i}u \tilde X_\theta}]\right| = o(\theta ^{\bar{\beta}+ \epsilon}),$$
    where $\bar{\beta} = \min\{1 + 2\alpha_\rho, \tfrac{3}{2}\}$ as defined in Lemma \ref{lemma 1}.
\end{lemma}
\begin{proof}
    From \textnormal{Eq.} \eqref{epsilon cond} and the fact that $\tilde{\sigma}_0(\theta) = O(\sqrt{\theta})$, we have $\|X_\theta - \tilde X_\theta\|_{1 + \epsilon} = o(\theta^{\bar{\beta} + 2\epsilon}).$ Then it follows from Assumption \ref{assump:EFGR} and the proof of Lemma 3.1  in \cite{euch2019short} by substituting $2H$ by $\bar{\beta}$ that the result follows.
\end{proof}
\begin{lemma}
\label{int diff2}
    We denote by $\tilde{M}_\theta^{(0)} = M_\theta^{(0)} - \tfrac{\tilde{\sigma}_0}{2}$. Given $\epsilon \in (0, \tfrac{1}{8})$ and $\alpha \in \mathbb{N} \cup \{0\}$, it holds that, for any $\delta \in [0, \tfrac{\bar{\beta}}{6} - \tfrac{\epsilon}{3})$, 
    \[
\sup_{|u| \le \theta^{-\delta}} \left| E\left[\tilde X_\theta^\alpha e^{\mathrm{i}u\tilde X_\theta} - e^{\mathrm{i}u\tilde{M}_\theta^{(0)}}\left((\tilde{M}_\theta^{(0)})^\alpha + A_\theta(\alpha, u, \tilde{M}_\theta^{(0)}) + B_\theta(\alpha, u, \tilde{M}_\theta^{(0)})\right)\right] \right| = o(\theta^{\bar{\beta}+\epsilon}),
\]
where
\[
A_\theta(\alpha, u, x) = (\mathrm{i}ux^\alpha + \alpha x^{\alpha-1}) \left( E[\tilde X_\theta \mid \tilde{M}_\theta^{(0)} = x] - x \right),
\]
\[
\begin{aligned}
B_\theta(\alpha, u, x) &= \left( -\tfrac{u^2}{2}x^\alpha + \mathrm{i}u \alpha x^{\alpha-1} + \tfrac{\alpha(\alpha-1)}{2}x^{\alpha-2} \right)\left(\theta E [ | M_\theta^{(1)} |^2 \mid \tilde{M}_\theta^{(0)} = x ]\right.\\
&\quad \quad \left.-\tilde{\sigma}_0\theta E[M_\theta^{(1)} M_\theta^{(3)} \mid \tilde{M}_\theta^{(0)} = x] \right).
\end{aligned}
\]
\end{lemma}
\begin{proof}
    By applying the trivial inequality 
    $$\left|e^{\mathrm{i}x} - 1 - \mathrm{i}x + \tfrac{x^2}{2}\right| \le \tfrac{|x|^3}{6}$$
    to $|x| = |\tilde X_\theta - \tilde{M}_\theta^{(0)}| = O(\theta^{\overline{\overline{\beta}}})$ with $\overline{\overline{\beta}} = \min\{\tfrac{1}{2} + \alpha_\rho, 1\}$, we have
    \[
\begin{aligned}
\sup_{|u|\le \theta^{-\delta}} \left| E\left[\tilde X_\theta^\alpha e^{\mathrm{i}u\tilde X_\theta}- \tilde X_\theta^\alpha e^{\mathrm{i}u\tilde{M}_\theta^{(0)}} \left(1 + \mathrm{i}u(\tilde X_\theta - \tilde{M}_\theta^{(0)}) - \tfrac{u^2}{2}(\tilde X_\theta - \tilde{M}_\theta^{(0)})^2 \right) \right] \right| = o(\theta^{\bar{\beta}+\epsilon}).
\end{aligned}
\]
    The order of the error term is due to $3\overline{\overline{\beta}} - 3\delta \ge \bar{\beta} + \epsilon$ by assumption. The result then follows from expanding $\tilde X_\theta^\alpha = (\tilde{M}_\theta^{(0)})^\alpha + \alpha (\tilde{M}_\theta^{(0)})^{\alpha - 1}(\tilde X_\theta - \tilde{M}_\theta^{(0)}) + \cdots$, and taking the conditional expectation with respect to $\tilde{M}_\theta^{(0)}$.
\end{proof}
\begin{lemma}
    Define \[
\begin{aligned}
\bar{q}_\theta(x) &= \phi(x+ \tfrac{\tilde{\sigma}_0}{2}) - \sqrt{\theta} a_\theta^{(1)}(x+ \tfrac{\tilde{\sigma}_0}{2}) - \theta a_\theta^{(2)}(x+ \tfrac{\tilde{\sigma}_0}{2}) \\
&\quad  \quad + \tfrac{\tilde{\sigma}_0}{2}\sqrt{\theta}a_\theta^{(3)}(x+ \tfrac{\tilde{\sigma}_0}{2}) + \tfrac{\theta}{2}c_\theta(x+ \tfrac{\tilde{\sigma}_0}{2}) - \tfrac{\tilde{\sigma}_0}{2}\theta b_\theta(x+ \tfrac{\tilde{\sigma}_0}{2}),
\end{aligned}
\] with $a_\theta^{(i)}$, $i = 1,2,3$, $b_\theta$ and $c_\theta(x)$ given in Lemma \ref{lemma 1}. Then
    $$\int_{\mathbb{R}}e^{\mathrm{i}ux}x^\alpha \bar{q}_\theta(x)\mathrm{d}x = E\left[e^{\mathrm{i}u\tilde{M}_\theta^{(0)}}\left(\left(\tilde{M}_\theta^{(0)}\right)^\alpha + A_\theta\left(\alpha, u, \tilde{M}_\theta^{(0)}\right) + B_\theta\left(\alpha, u, \tilde{M}_\theta^{(0)}\right)\right)\right].$$
\end{lemma}
\begin{proof}
    This directly follows from integration by parts.
\end{proof}
\begin{lemma}
    For any $\epsilon \in (0, \tfrac{1}{8})$, we have 
    $$\begin{aligned}
\sup_{x\in\mathbb{R}} (1+x^2)^\alpha |q^{(1,2)}_\theta(x) - \bar{q}_\theta(x)| = o(\theta^{\bar{\beta}}),
\end{aligned}$$
    where $q^{(1,2)}_\theta(x)$ is given by \textnormal{Eq.} \eqref{q} with $\beta_1 = \tfrac{1}{2} + \alpha_\rho$, $\beta_2 = 1$ and $\kappa_3(\theta)$, $\kappa_4(\theta)$ are given by \textnormal{Eq.} \eqref{kappa34}.
\end{lemma}
\begin{proof}
    From \textnormal{Eq.} \eqref{abc val} and \textnormal{Eq.} \eqref{b}, we have
    $$\begin{aligned}\bar{q}_\theta(x) &= \phi(x + \tfrac{\tilde{\sigma}_0}{2})\left(1 + \kappa_3\left(H_3(x+\tfrac{\tilde{\sigma}_0}{2}) - \sigma_0H_2(x+\tfrac{\tilde{\sigma}_0}{2})\right)\theta^{\beta_1}\right)\\
    &\quad + \phi(x+\tfrac{\tilde{\sigma}_0}{2})\kappa_4\theta^{\beta_2}\left(H_4(x+\tfrac{\tilde{\sigma}_0}{2}) - 2 \tilde{\sigma}_0(\theta) H_3(x+ \tfrac{\tilde{\sigma}_0}{2})\right)\\
    &\quad + \phi(x+\tfrac{\tilde{\sigma}_0}{2})\tfrac{\kappa_3^2}{2} H_6\left(x+\tfrac{\tilde{\sigma}_0}{2}\right) \theta^{2\beta_1}.
    \end{aligned}$$
    Next, from the proof of Proposition \ref{substitute}, we have
    $$\begin{aligned}|\tilde{\sigma}_0^2 - \sigma_0^2| &\le \tfrac{1}{4}\operatorname{Var}\left(\int_0^\theta v_t\mathrm{d}t\right) + \left|E\left[\int_0^\theta \sqrt{v_t}\mathrm{d}B_t \cdot \int_0^\theta v_t\mathrm{d}t\right]\right|\\
    &= O(\theta^3) + \left|E\left[\int_0^\theta \sqrt{v_t}\mathrm{d}B_t \cdot \int_0^\theta \left(\int_0^t \mathcal{L}v(Y_s)\mathrm{d}s + \int_0^t v^\prime(Y_s)c(Y_s)\mathrm{d}W_s\right)\mathrm{d}t\right]\right|\\
    & = O(\theta^{\tfrac{5}{2}}) + O(\theta^{2 + \alpha_\rho})\\
    &= o(\theta^{\bar{\beta} + \tfrac{1}{2}}),\end{aligned}$$
    where the last expression is due to $\min\{\tfrac{3}{2} + 2\alpha_\rho, 2\} < \min\{\tfrac{5}{2}, 2+ \alpha_\rho\}$. Thus, combined with the definitions of $\bar q_\theta$ and $q_\theta^{(1,2)}$, we have
    $$|\bar{q}_\theta(x) - q^{(1,2)}_\theta(x)| = O(\sigma_0 - \tilde{\sigma}_0)= o(\theta^{\bar \beta}).$$
    The result follows by observing that $\tilde{q}_\theta(x)$ and $q_\theta^{(1,2)}(x)$ are Schwartz functions.
\end{proof}

It remains to show that the density $p_\theta(x)$ exists and 
$$\sup_{x\in \mathbb{R}}(1 + x^2)^\alpha |p_\theta(x) -\bar{q}_\theta(x)| = o(\theta^{\bar{\beta}}),$$
which, under Assumption~\ref{assump:EFGR}, follow from the proof of Theorem 2.1 in \cite{euch2019short} by substituting $2H$ by $\bar\beta$ and combining the results of Lemma \ref{int diff} and Lemma \ref{int diff2}. Finally, we have shown that condition \eqref{density convergence} is satisfied by $q^{(1,2)}_\theta$, $\beta_1 = \tfrac{1}{2} + \alpha_\rho$, $\beta_2 = 1$, and the rest of the results in Proposition \ref{decaying leverage} follow from Theorem \ref{Th1} and Theorem \ref{ATM skew} by taking $\bar \beta_\epsilon = \bar \beta$.

\subsection{Details of Empirical Asymptotics Analysis}
\label{data process}
The empirical asymptotics analysis follows the methodology outlined in \cite{glasserman2020buy}. The dataset comprises S\&P 500 index option quotes from January 3, 2023, to February 28, 2023, including the implied volatility computed from CBOE. For each expiry $\theta$, the forward price $F(\theta)$ is recovered from put--call parity using the call and put quotes closest to the at-the-money strike $K$:
\[
F(\theta) = K + e^{r(\theta)\theta}\big(C(K,\theta) - P(K,\theta)\big),
\]
where $r(\theta)$ is the prevailing risk-free rate at maturity $\theta$, obtained by linearly interpolating the daily Treasury par yield curve rates to the option maturity $\theta$. The standardized log-moneyness is then defined as $z = \log(K/F(\theta))/\sqrt{\theta}$. The dataset is filtered according to the following criteria:
\begin{itemize}[itemsep=1pt]
    \item Only option quotes with positive open interest and positive trading volume are included.
    \item Only out-of-money options are included, i.e., call options with moneyness $> 1$ and put options with moneyness $ < 1$.
    \item Time to maturity must range from 3 days to 90 days.
    \item All observations must have non-missing implied volatility.
    \item Only options with the ``SPX" symbol are included (with SPX weekly options excluded).
    \item Standardized moneyness $z$ must be within $[-0.75, 0.75]$.
\end{itemize}

To mitigate the sensitivity of ATM derivatives to short-term curvature and bid-ask noise, we fit a cubic smoothing spline to the implied volatility points for every maturity. To ensure the fit performs with sufficient accuracy, further conditions are imposed within each maturity bucket:
\begin{itemize}[itemsep=1pt]
    \item The maximum standardized moneyness must be at least 0.5.
   \item At least four implied volatility observations are required for both out-of-the-money puts and out-of-the-money calls.
   \item Outliers (i.e., with clear violation of arbitrage-free condition) are removed.
\end{itemize}

Finally, we use a power-law function to fit the ATM skew and curvature across different maturities that are calculated from the coefficients of the fitted splines. For the power-law fitting procedure, a log transformation is applied, followed by linear regression to enhance robustness. This step mandates the presence of at least four distinct maturities. Furthermore, the first three skews or curvatures must exhibit a monotonic ordering to ensure the consistency of the fitted relationship. Otherwise, this set of implied volatility points is interpreted as not robust and discarded.

\bibliographystyle{rQUF}
\bibliography{rQUFguide}

\end{document}